\documentclass[useAMS,usenatbib]{mnras}
\usepackage{graphicx,natbib,amssymb,amsmath,stmaryrd,makecell,placeins}
\usepackage{soul}
\usepackage{float}
\usepackage{dblfloatfix}
\usepackage{subcaption}
\usepackage{microtype}
\usepackage{textcase}
\usepackage[dvipsnames]{xcolor}
\usepackage{fancyhdr}
\usepackage{multirow}
\usepackage{rotating}
\begin{document}

\title[Including massive neutrinos in SZ  power spectrum and cluster counts analyses]{Including massive neutrinos in thermal Sunyaev Zeldovich  power spectrum and cluster counts analyses}

\author[B. Bolliet et al.]{Boris Bolliet$^{1}$\thanks{boris.bolliet@gmail.com}, Thejs Brinckmann$^{2,3}$\thanks{thejs.brinckmann@stonybrook.edu}, Jens Chluba$^{1}$, Julien Lesgourgues$^{3}$\\
$^{1}$Jodrell Bank Centre for Astrophysics, School of Physics and Astronomy, The University of Manchester, Manchester, M13 9PL, U.K.\\
$^{2}$C.N. Yang Institute for Theoretical Physics and Department of Physics \& Astronomy,
Stony Brook University,\\ Stony Brook, NY 11794, USA\\
$^{3}$Institute for Theoretical Particle Physics and Cosmology (TTK), RWTH Aachen University, D-52056 Aachen, Germany\\
}
\maketitle
\thispagestyle{fancy}
\rhead{\small Preprint TTK-19-24 \& YITP-SB-19-19}
\vspace{-45mm}
\begin{abstract}
We consistently include the effect of massive neutrinos in the thermal Sunyaev Zeldovich (SZ) power spectrum and cluster counts analyses, highlighting subtle dependencies on the total neutrino mass and data combination. In particular, we find that using the transfer functions for Cold Dark Matter (CDM) + baryons in the computation of the halo mass function, instead of the transfer functions including neutrino perturbations, as prescribed in recent work, yields a $\approx$ 0.25\% downward shift of the $\sigma_8$ constraint from tSZ power spectrum data, with a fiducial neutrino mass $\Sigma m_\nu=0.06$ eV. In $\Lambda$CDM, with an X-ray mass bias corresponding to the expected hydrostatic mass bias, i.e., $(1-b)\simeq0.8$, our constraints from Planck SZ data are consistent with the latest results from SPT, DES-Y1 and KiDS+VIKING-450. In $\nu\Lambda$CDM, our joint analyses of Planck SZ with Planck 2015 primary CMB yield a small improvement on the total neutrino mass bound compared to the Planck 2015 primary CMB constraint, as well as $(1-b)=0.64\pm0.04$~(68\%~CL). For forecasts, we find that competitive neutrino mass measurements using cosmic variance limited SZ power spectrum require masking the heaviest clusters and probing the small-scale SZ power spectrum up to  $\ell_\mathrm{max}\approx10^4$. Although this is challenging, we find that SZ power spectrum can realistically be used to tightly constrain intra-cluster medium properties: we forecast a 2\% determination of the X-ray mass bias by combining CMB-S4 and our mock SZ power spectrum  with $\ell_\mathrm{max}=10^3$.  
\end{abstract}


\section{Introduction}
Observationally, the structure of the present Universe appears as a web of cosmic voids and dense filaments filled with baryonic matter and a cold dark matter (CDM). At the crossroads of these filaments lie clusters of several hundreds of galaxies, constituting the largest known gravitationally bound objects. In the past decades, the properties of galaxy clusters have been determined using several techniques including weak lensing, optical, X-ray and Sunyaev Zeldovich (SZ) surveys \citep[see, e.g.,][]{Burenin:2006td,Vikhlinin:2008cd,Ade:2015gva,2016ApJ...832...95D,Hilton:2017gal,2018A&A...615A.112R}.  Galaxy clusters have a typical mass of a few $10^{14} \mathrm{M}_\varodot$. About ten percents of the mass is a hot electron gas with temperatures reaching $T\approx 5$$-$$10\,\mathrm{keV}$, while galaxies themselves are responsible for only about one percent of the total mass. The remainder of the mass is made of CDM.  On scales smaller than the cluster size, observations and simulations have shown that the structure of the intra-cluster medium (ICM)  is well described by a Navarro-Frenk-White (NFW) profile for the CDM density \citep{Navarro:1995iw} that translates into a generalised NFW pressure profile for the hot gas  of electrons \citep{2007ApJ...668....1N,2010A&A...517A..92A}. Since galaxy clusters form in the potential wells created by CDM, their large-scale distribution directly traces the underlying cosmological perturbation field. Therefore, their abundance and spatial distribution provide a powerful cosmological probe and it is crucial to understand how subtle effects, e.g., caused by massive neutrinos, propagate into cluster observables. 

Neutrino oscillations experiments show that neutrinos are massive \citep[see, e.g.,][]{Kajita:2016cak,deSalas:2017kay,Esteban:2018azc}. The three mass states can be arranged in the so-called normal hierarchy, with one mass state significantly larger than the other two, or the inverted hierarchy, where two mass states are significantly larger than the third. Respectively, these configurations imply  $\Sigma m_{\nu} \gtrsim 0.06\,\mathrm{eV}$ and $\Sigma m_{\nu} \gtrsim 0.1\,\mathrm{eV}$ for the total neutrino mass. Current constraints from cosmology give $\Sigma m_{\nu} \lesssim 0.12\,\mathrm{eV}$ at 95\% CL~\citep[e.g.,][]{Palanque-Delabrouille:2015pga,Cuesta:2015iho,Giusarma:2016phn,Vagnozzi:2017ovm,Aghanim:2018eyx} combining Cosmic Microwave Background (CMB) and Baryon Acoustic Oscillations (BAO) data in $\Lambda$CDM. This translates into upper bounds on the individual neutrino mass states, but the hierarchy is still unknown.

Significant progress have been made in recent years towards understanding neutrino mass effects in cosmology in order to improve upon current constraints~\citep[e.g.,][]{Allison:2015qca,Liu:2015txa,LoVerde:2016ahu,Archidiacono:2016lnv,Alam:2016hwk,Lorenz:2017fgo,Boyle:2017lzt,Yu:2018tem,Boyle:2018rva,Oldengott:2019lke}. A neutrino mass sum measurement is furthermore expected from future large-scale structure and galaxy surveys such as DESI, LSST, Euclid and SKA~\citep[e.g.][]{Audren:2012vy,Villaescusa-Navarro:2015cca,Calabrese:2016eii,DiValentino:2016foa,Obuljen:2017jiy,Sprenger:2018tdb,Mishra-Sharma:2018ykh,Brinckmann:2018owf}. Aside from determining the neutrino mass hierarchy, a neutrino mass sum detection from cosmology would put lower bounds on the two larger mass states and, in the case of a non-minimal normal hierarchy, a lower bound on the lightest neutrino mass state.\footnote{For reviews on neutrino cosmology, see, e.g., ~\citet{Lesgourgues:2006nd,Hannestad:2010kz,Lesgourgues:2018ncw,Abazajian:2016yjj,Lattanzi:2017ubx} or more foundational articles such as \citet{Hu:1997mj,Bashinsky:2003tk}.}

The thermal SZ power spectrum probes galaxy clusters on cosmological scales via the SZ effect, a spectral distortion of the CMB radiation spectrum due to the up-scattering of CMB photons by the hot electrons of the ICM. The SZ effect was predicted by \cite{1969Ap&SS...4..301Z} and \cite{1972CoASP...4..173S} and first detected by \cite{Birkinshaw:1984aa}.  \cite{Komatsu:1999ev} provided the formalism for the accurate calculation of the SZ power spectrum and advocated its use in the cosmological context, as further developed by \cite{Komatsu:2002wc}.\footnote{For  reviews on SZ science, see, e.g., \cite{Carlstrom:2002na} and \cite{Mroczkowski:2018nrv}, where the latter highlights the potential of future high-resolution SZ observations.} 
These previous works developed the tools to use the SZ power spectrum to shed light on standard cosmological parameters, in particular $\sigma_8$ and $\Omega_\mathrm{m}$. The potential for SZ observables to probe extended cosmologies, e.g., related to primordial non-Gaussianity, massive neutrinos and dark energy, was emphasised by, e.g.,  \cite{Weinberg:2012es, Hill:2013baa, Roncarelli:2014jla,Ade:2015fva,McCarthy:2017csu, Salvati:2017rsn,Bolliet:2017lha,Ade:2018sbj,Bocquet:2018ukq}. In addition, the evolution of the number of clusters, i.e., `cluster counts',  with redshift has long been known to provide a sensitive probe of cosmology \citep[e.g.,][]{1986MNRAS.222..323K, 1991ApJ...383..104K, 1989ApJ...347..563P, Barbosa1996, Holder2001,Battye2003}. Here we follow \cite{Costanzi:2013bhatex} for the calculation of the halo mass function (HMF) when neutrinos are massive, and extend the framework developed for \verb|class_sz| \citep{Bolliet:2017lha} to consistently include massive neutrinos in the theoretical models for the the SZ power spectrum and cluster counts.

Recent experiments have produced large catalogues of cluster SZ observations, in particular Atacama Cosmology Telescope \citep[ACT,][]{Hasselfield:2013wf}, South Pole Telescope \citep[SPT,][]{Bleem:2014iim} and the Planck satellite \citep{Ade:2015gva}. Moreover, the Planck Collaboration obtained the first all-sky Compton-$y$ map and its associated SZ power spectrum  \citep{Ade:2013qta}. From these data sets, a number of works have performed cosmological parameter extraction analyses \citep[e.g.,][]{Hill:2014pyr,Aghanim:2015eva,Ade:2015fva, Hurier:2017jgi,Salvati:2017rsn,Bolliet:2017lha,Bocquet:2018ukq}.
Most of these analyses have obtained a best-fit cosmology with a best-fit matter clustering amplitude $\sigma_8$ noticeably lower than the one from Planck 2015 primary CMB data  (by $\approx 1-2\sigma$). This motivated a stream of research that attempted to identify the origin of the mismatch \citep[e.g.,][]{Dolag:2015dta,Bocquet:2015pva,Horowitz:2016dwk,Bolliet:2017lha,Makiya:2018pda,Remazeilles:2018laq,Salvati:2019zdp,Ruppin:2019deo}. In particular, for analyses based on the X-ray calibrated $Y_\mathrm{SZ}\textendash M$ scaling relations \citep[see, e.g.,][]{Nagai:2005wx}, a significant attention has been brought to the hydrostatic mass bias (see Section~\ref{ref:ssb} for a discussion on the mass bias).

Our motivations in this article are: to consistently include and study the effects of massive neutrinos on the SZ power spectrum and cluster counts (Section~\ref{sec:model}-\ref{s:pheno}); to update the constraints on matter clustering, X-ray mass bias and total neutrino mass coming from the Planck SZ data and compare with other cluster and galaxy survey constraints (Section~\ref{ss:pc}-\ref{ss:SZcomp-b}); to discuss whether massive neutrinos or dark energy can help explaining why the X-ray mass bias extracted from the combination  of Planck SZ plus Planck  primary CMB is in tension with the value expected by standard departure from HSE  (Section~\ref{ss:szcmb}-\ref{ss:wcdm}); to assess the constraining power of future SZ power spectrum measurements (Section~\ref{s:f}). 

In Section~\ref{sec:model} we present the calculation of the HMF and the SZ power spectrum for cosmological models with massive neutrinos. We then discuss the importance of the `cb' prescription for the HMF in Section~\ref{s:cb}. Section~\ref{s:pheno} reviews the effects of massive neutrinos on the SZ power spectrum and cluster counts, while in Section~\ref{sec:results} we present our constraints on the total neutrino mass based on the Planck SZ data. In Section~\ref{s:f} we present forecasts for cosmic variance limited SZ power spectrum experiments. We discuss our results and conclude in Section~\ref{sec:conclusion}. Our constraints are summarised in tables \ref{ref:table-obs} and \ref{ref:table-forecast} in Sections \ref{sec:results} and \ref{s:f}.

\subsection*{Definitions and general settings}\label{sec:def}
We study three degenerate active neutrinos\footnote{ The degenerate scenario is a better approximation to the realistic normal hierarchy and inverted hierarchy scenarios than models with one massive plus two massless or one massless plus two massive species~\citep{Lesgourgues:2006nd}. }, with a total neutrino mass $\Sigma m_\nu$, in spatially flat $\Lambda$CDM or $w$CDM cosmologies with constant dark energy equation of state $w$. We keep the effective number of neutrino species fixed to the nominal value $N_\mathrm{eff} = 3.046$ \citep[e.g.,][]{Mangano:2005cc,deSalas:2016ztq}.

We use the notations $n_\mathrm{s}$ and $A_\mathrm{s}$ for the spectral index and the amplitude of the primordial power spectrum of curvature perturbation; $\rho_\mathrm{crit}$ for the critical density of the universe today; $\Omega_\mathrm{c}$, $\Omega_\mathrm{b}$, $\Omega_\gamma$, $\Omega_\nu$ for the present CDM, baryons, photons and neutrino density fractions respectively; $\Omega_\mathrm{m} = \Omega_\mathrm{c} + \Omega_\mathrm{b} + \Omega_\nu+ \Omega_\gamma$ for the present matter density fraction; $\tau$ for the reionization optical depth;  $H_0$ and $h=H_0/(100\,\mathrm{km/s/Mpc})$ for the Hubble constant and reduced Hubble constant; $\theta_\mathrm{s}$ for the angular size of the sound horizon at decoupling;  $f_\nu\equiv\Omega_\nu/\Omega_\mathrm{m}$ for the neutrino fraction; $\omega_x = \Omega_x h^2$ for the physical densities of  species $x=\mathrm{c,b,\nu,m}$; $\sigma_8$ for the variance of the matter over-density field smoothed by a sphere of radius $8\,\mathrm{Mpc}/h$.

For our numerical calculations, unless otherwise stated, we use a fiducial (or reference) $\Lambda$CDM cosmology with parameters: $h=0.7$, $\Omega_\mathrm{b} = 0.05$, $\Omega_\mathrm{c}=0.26$, $\tau=0.07$, $A_\mathrm{s}=1.93\times 10^{-9}$, $n_\mathrm{s}=0.96$ and $\Sigma m_\nu =0.06\,\mathrm{eV}$. This model has $\sigma_8 = 0.80$ and $100\theta_\mathrm{s}=1.0510$. 
For the modelling of the SZ power spectrum and cluster counts, we use the \cite{Tinker:2008ff} HMF interpolated at $M_{500c}$, the \cite{2013A&A...550A.131P} pressure profile  \citep[see][for details]{Bolliet:2017lha}, and, when not varied or unless otherwise stated, a reference X-ray mass bias $B=1.41$ corresponding to $(1-b)=1/B\simeq 0.7$, i.e. a bias of 30\%. We denote as $B_\mathrm{HSE}=1.25$, i.e.,  $(1-b_\mathrm{HSE})=0.8$, the values associated with standard departure from hydrostatistical equilibrium in the ICM \citep[e.g.,][]{Kay:2004wp,2007ApJ...655...98N,Shi:2015fua}. For the tSZ power spectrum computation we integrate over masses between $10^{11}\mathrm{M}_\varodot/h$ and $5\times10^{15}\mathrm{M}_\varodot/h$, and over redshifts between $z=0$ and $z=3$.

\section{SZ power spectrum and cluster counts with massive neutrinos}\label{sec:model}
Here we recall the theoretical framework for the computations of the SZ power spectrum (Section~\ref{ss:ps}) and SZ cluster counts (Section~\ref{ss:cc}) with massive neutrinos. Then, in Section~\ref{ref:ssb} we briefly review the role the X-ray mass bias parameter in the Planck SZ analyses.
\subsection{Power spectrum}\label{ss:ps}

We compute the SZ power spectrum according to the halo model of structure formation (see \cite{Cooray:2002dia} for a review of the halo model, and \cite{Komatsu:1999ev} for its application to the SZ power spectrum). Here, we only consider the $1$-halo contribution to the SZ power spectrum, as the 2-halo term is at least one order of magnitude smaller on scales of observational interest \citep[e.g.,][]{Komatsu:1999ev, Hill:2013baa, Horowitz:2016dwk}.  The model of the SZ power spectrum has three main ingredients: the differential volume element, $\mathrm{d}V/(\mathrm{d}z\mathrm{d}\Omega)$ with respect to redshift and solid angle; the HMF, $\mathrm{d}N/\mathrm{d}M\mathrm{d}V$; and the square of the 2d Fourier transform of the ICM electron pressure profile integrated over the line-of-sight, $|y_\ell (M,z)|^2$.  The angular power spectrum of the SZ effect is obtained by integrating the product of these three quantities over redshift, $z$, and cluster masses, $M$:
\begin{equation}
C_\ell^{\mathrm{tSZ}} = \int  \mathrm{d}z\int \mathrm{d}M \frac{\mathrm{d}V}{\mathrm{d}z\mathrm{d}\Omega}\frac{\mathrm{d}N}{\mathrm{d} M \mathrm{d}V} |y_\ell (M,z)|^2.\label{eq:tSZPS}
\end{equation}
With massive neutrinos, the differential volume element and Fourier transform of the pressure profile can be computed without modifications \citep[see, e.g., ][for details about the computation of the volume element, the HMF and the pressure profile]{Bolliet:2017lha}. Indeed, since neutrinos constitute a tiny fraction of the halo masses and do not interact with baryons, they are expected to have negligible effects on the electron pressure profiles. However, the calculation of the HMF needs some revision due to the presence of massive neutrinos.  \cite{Costanzi:2013bhatex} and \cite{Castorina:2013wga} performed N-body simulations incorporating massive neutrinos. In agreement with \cite{Ichiki:2011ue} and subsequent work of \cite{LoVerde:2014rxa}, they showed that the \cite{Tinker:2008ff} fitting formula remains valid for the HMF, provided one does not use the total linear matter power spectrum and the total matter density but rather the power spectrum of CDM and baryons, $P_\mathrm{cb}(k,z)$, and the density $\rho_\mathrm{cb}=(\Omega_\mathrm{c} + \Omega_\mathrm{b})\rho_\mathrm{crit}$ in the calculation of the variance of the smoothed over-density field $\sigma(M,z)$. More explicitly, the square of this quantity is given by  
\begin{equation}
\sigma^2 (M,z) = \int \frac{\mathrm{d}k}{k} \frac{k^3}{2\pi^2} P_{\mathrm{cb}}(k,z) W^2 (kR), \label{eq:sigma}
\end{equation}
where $R=[3M/4\pi\rho_\mathrm{cb}]$ and $W(x)$ is the top-hat window function.
This `cb'  prescription is consistent with the physical expectation that neutrinos do not cluster as much as baryons and CDM, and thus, do not trigger the formation of compact objects \citep[see, e.g.,][for details]{LoVerde:2014rxa}. 

\subsection{Cluster counts}\label{ss:cc}
%
SZ cluster surveys are sensitive to the cluster signal-to-noise. For a cluster of mass $M$ at redshift $z$, the predicted signal-to-noise is computed using the $Y_\mathrm{SZ}-M$ and $\theta-M$ relation, where $Y_\mathrm{SZ}$ is the SZ flux and $\theta$ the angular size associated with the cluster. We use the same relations as \cite{Ade:2015fva}, where the spherical over-density mass defined with respect to $500\rho_\mathrm{crit}$, say $M^\mathrm{X\textendash ray}_{500\mathrm{c}}$, is a biased estimator of the true halo mass $M^\mathrm{true}_{500\mathrm{c}}$, i.e., $M^\mathrm{X\textendash ray}_{500\mathrm{c}}=M^\mathrm{true}_{500\mathrm{c}}/B$, with $B$ the X-ray mass bias discussed in the next subsection. Once $\theta$ and $Y_\mathrm{SZ}$ are known for a given mass, we compute the signal-to-noise as $\xi = Y_\mathrm{SZ}(M,z)/\sigma(\theta;l,b)$
where $\sigma$ is obtained by interpolating the experiment's noise map at galactic coordinate\footnote{Here $b$ is a galactic coordinate and has nothing to do with the X-ray mass bias also dubbed $b$ in the next Section.} $(l,b)$ and angular size $\theta$. 
The theoretical number of clusters in each signal-to-noise bin centred on $\xi_j$ and redshift bin centred on $z_i$  is then given by $\bar{N}_{ij}=(\mathrm{d}N/\mathrm{d}z\mathrm{d}\xi) \Delta z_i \Delta \xi_j$, with
\begin{equation}
\frac{\mathrm{d}N}{\mathrm{d}z\mathrm{d}\xi}= \int  \mathrm{d}\Omega\int \mathrm{d}M \frac{\mathrm{d}V}{\mathrm{d}z\mathrm{d}\Omega}\frac{\mathrm{d}N}{\mathrm{d} M \mathrm{d}V}\mathcal{P}(\xi,\xi_j). \label{eq:dndzdq}
\end{equation}
Here $\mathcal{P}$ is the detection probability  for clusters in the signal-to-noise bin $\xi_j$.  For the detection probability we use an analytical approximation with the `erf' completeness \citep[see Eq. 14 in][]{Ade:2015fva}, and  a detection threshold that we keep fixed to $\xi_\mathrm{cut}=6$, as was done in the baseline \cite{Ade:2015fva} analysis. For redshift and mass range to compute the predicted cluster counts, we also used the settings of the Planck likelihood code \verb|szcounts.f90| which has $z_\mathrm{max}=1$ and $M_\mathrm{min}=5\times 10^{13}\,\mathrm{M}_\varodot/h$, $M_\mathrm{max}=1.2\times 10^{16}\,\mathrm{M}_\varodot/h$.

\subsection{Role of the X-ray mass bias parameter}\label{ref:ssb}
A mass bias parameter is present in the Planck SZ analyses because the scaling relations between the SZ flux and size of clusters on one hand, and the cluster masses, on the other hand, are calibrated on X-ray measurements that use hydrostatic mass estimates \citep[see Appendix A of][]{Ade:2013lmv}. The `true' mass  of a cluster is not a direct observable. It can only be inferred from the measurements of other observables, that constitute \textit{mass proxies}, complemented by assumptions for the hydrodynamical state and geometry of the cluster. \cite{Kravtsov:2006db} discussed several mass proxies, for instance the X-ray temperature $T_\mathrm{X}$, X-ray luminosity $L_\mathrm{X}$, the integrated SZ flux of the ICM $Y_\mathrm{SZ}$ \citep[e.g.,][]{Nagai:2005wx}, or  the low-scatter $Y_\mathrm{X}\equiv M_\mathrm{gas}T_\mathrm{X}$ introduced by \cite{Kravtsov:2006db}. One then relies on models and assumptions to compute the cluster mass from X-ray observations. For instance, the masses of the Chandra sample of \cite{Vikhlinin:2005mp} are estimated using models for the gas density and temperature profiles (calibrated on X-ray observations) as well as the assumption of hydrostatic equilibrium (HSE) and spherical symmetry \citep[e.g.,][]{RevModPhys.58.1}. Then it is possible to construct the relation between the mass proxy, $Y_\mathrm{X}$ or $Y_\mathrm{SZ}$, and the X-ray mass estimate, or eventually the spherical over-density mass $M^\mathrm{X\textendash ray}_{500\mathrm{c}}$, by fitting the observational data.\footnote{Since the mass estimate that enters the constructed $Y_\mathrm{X}-M^\mathrm{X\textendash ray}_{500\mathrm{c}}$ or $Y_\mathrm{SZ}-M^\mathrm{X\textendash ray}_{500\mathrm{c}}$ relations relies on HSE, it is some times called the hydrostatic mass.} 

The true mass is generally larger than the HSE mass. Numerical simulations suggest that the HSE mass is biased low by $ 20\%$ with respect to the true mass, corresponding to $B_\mathrm{HSE}=1.25$ \citep[e.g.,][]{Kay:2004wp,2007ApJ...655...98N,Lau:2009qm,Battaglia2012, Nelson:2013hwa}. Such bias arises due to non-thermal pressure and recent analytical work have been successful in modelling it fairly \citep[e.g.,][]{Shi:2014msa,Shi:2014lua,Shi:2015fua}. Although numerical work relying on the  ART code  \citep{2002ApJ...571..563K} find a HSE bias that is  mass independent, the recent BAHAMAS and MACSIS simulations are suggesting a mass dependent HSE bias that reaches $\approx40\%$ for heavy clusters with $M_\mathrm{500c}\gtrsim 10^{15}\mathrm{M_\varodot}$ \citep{Henson:2016eip}. In this article we do not study mass-dependent bias, however, we note that a larger bias for heavier clusters would have the effect of boosting the SZ power at large scales where heavier clusters contribute more \citep[e.g., Figure 1 of][]{Remazeilles:2018laq}. This could drive the value of $\sigma_8$ extracted from the SZ power spectrum analysis to a larger value, more consistent with Planck primary CMB.

 In fact, the X-ray mass bias can be used as a phenomenological parameterisation of departure from other assumptions than HSE,   and of systematics and physical processes that are yet to be better understood \citep[e.g.,][]{Pratt:2019cnf}. Here, we either use the mass bias as a free parameter, or set a Gaussian prior around the typical value expected   from numerical hydrodynamical simulations, i.e., $B_\mathrm{HSE}=1.25$ or $(1-b_\mathrm{HSE})=0.8$ \citep[e.g.,][]{Kay:2004wp,2007ApJ...655...98N,Shi:2015fua}. Several observational constraints on the X-ray mass bias are shown on the right panel of Figure~\ref{fig:comps8}, along with the results we obtain in Section \ref{sec:results}. The relative precision on the X-ray mass bias is  currently at the $\sim10\%$ level and is expected to reach the 1\% level in the next decade \citep[e.g.,][]{Louis:2016gvv}. 

\section{Importance of the `cb' prescription for tSZ power spectrum}\label{s:cb}

\begin{figure}
\begin{centering}
\includegraphics[height=6cm]{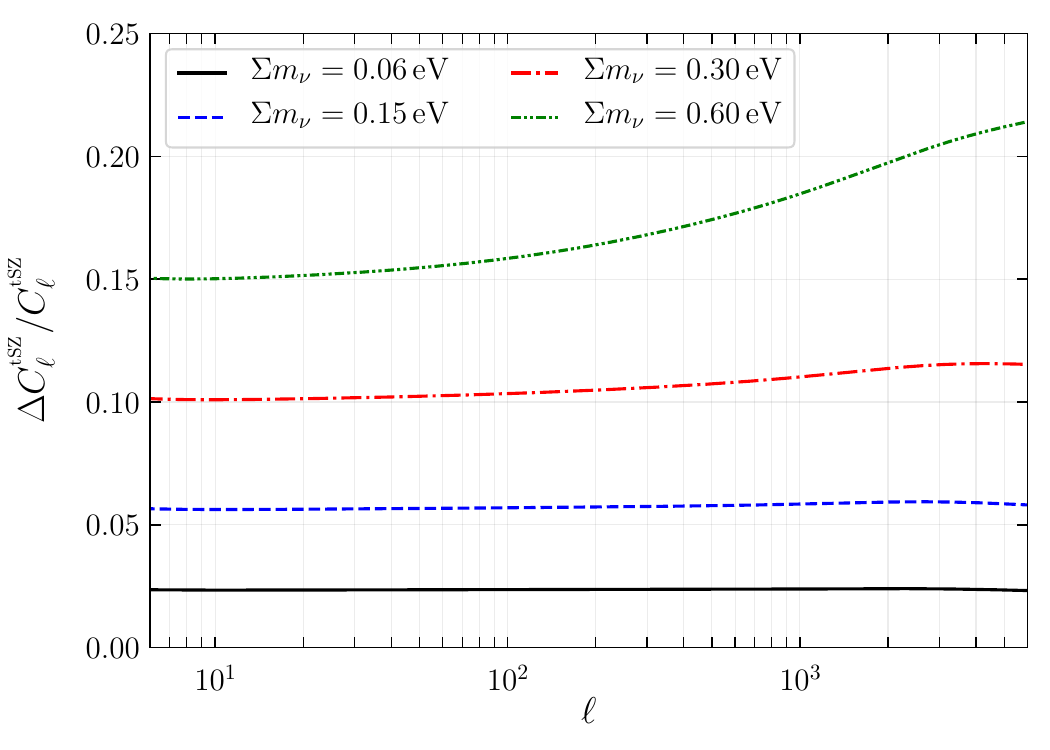}
\par\end{centering}
\caption[]{
Relative difference between SZ power spectrum computed with and without the `cb' prescription.
We use the notation $\Delta Q/Q \equiv (Q^\mathrm{cb}-Q)/Q$ where the superscript `cb' indicates the correct modelling. For instance, at $\Sigma m_{\nu}=0.30\,\mathrm{eV}$, omitting  the `cb' prescription leads to an under-estimation of the SZ power spectrum amplitude by about 10\%. Here, $\Sigma m_\nu$ is varied while  ($A_\mathrm{s}$, $n_\mathrm{s}$, $\omega_\mathrm{c}$, $\omega_\mathrm{b}$, $\theta_\mathrm{s}$) are set to their fiducial values: this choice is `case B' in section \ref{s:pheno}.  
}
\label{fig:pres1}
\end{figure}

For the numerical computations of the SZ power spectrum and  SZ cluster counts, we use \verb|class_sz|\footnote{\href{https://github.com/borisbolliet/class_sz}{https://github.com/borisbolliet/class\_sz}} \citep{Bolliet:2017lha}, a publicly available extension of the \verb|class| code~\citep{2011JCAP...07..034B}. We updated \verb|class_sz| to take into account the `cb' prescription. \cite{Ichiki:2011ue}, \cite{Costanzi:2013bhatex}, \cite{Castorina:2013wga} and \cite{LoVerde:2014rxa} have studied the effect of the `cb' prescription on the halo mass function and cluster counts. Quantitatively the authors of \cite{Castorina:2013wga} reported that while the `cb' prescription reproduces the results of N-body simulations for the halo abundance within 10\%, not using the `cb' prescription can lead to differences larger than 10-30\%, especially for the most massive haloes ($M\gtrsim 10^{15} \mathrm{M_{sun}}/h$) and/or at high redshift $z\approx 1$ (see their Figure 1). In \cite{Costanzi:2013bhatex} the authors showed that this difference in the HMF propagates to cluster counts in a way that can bias the resulting constraints on $\sigma_8$: not using the `cb' prescription yields a $\sigma_8$ biased high by $\approx 2\%$.
 Here we  present the effects of using the `cb' prescription on the tSZ power spectrum. Figure~\ref{fig:pres1} shows the relative difference for the SZ power  spectrum that arises when the `cb' prescription is used, for several values of $\Sigma m_\nu$. This difference can be understood as follows:
the current fraction of non-relativistic matter density in the form of neutrinos, $f_\nu=\omega_\nu / \omega_\mathrm{m}$, is given by $f_\nu \simeq \Sigma m_\nu / 14 \, \mathrm{eV}$ with our fiducial parameter values. The total matter power spectrum is given by a weighted sum over the `cb' power spectrum, the neutrino one and the cross-correlation term: $P_\mathrm{m} = (1-f_\nu)^2 P_\mathrm{cb} + f_\nu^2 P_\nu + 2 f_\nu (1-f_\nu) P_{\mathrm{cb}\times\nu}$. On scales smaller than the neutrino free-streaming scale, neutrino density perturbations are much smaller than baryon and CDM perturbations, such that $P_\mathrm{m} \simeq (1-f_\nu)^2 P_\mathrm{cb}$. Since this is true in particular for the scale $R=8\, h^{-1}\mathrm{Mpc}$, the relation  between the corresponding values of $\sigma_8$ is
\begin{equation}
\sigma_8^\mathrm{m} \simeq (1-f_\nu) \sigma_8^\mathrm{cb}~,
\label{eq:sigma8}
\end{equation}
where we used superscript `m'  and `cb' to indicate wether $\sigma_8$ is computed with $P_\mathrm{m}$ or $P_\mathrm{cb}$ .
Next, we recall that the SZ power spectrum scales as $(\sigma_8^\mathrm{m})^{8.1}\Omega_\mathrm{m}^{3.2}$ (for $\ell<10^3$) without 
massive neutrinos. According to the `cb' prescription, it now scales as $(\sigma_8^{\mathrm{cb}})^{8.1}\Omega_\mathrm{cb}^{3.2}$ in presence of massive neutrinos. Using Eq.~(\ref{eq:sigma8}) and $\Omega_\mathrm{cb} = (1-f_\nu) \Omega_\mathrm{m}$, we deduce that the relative difference between both should be given by $\Delta C_\ell^{\mathrm{tSZ}}/C_\ell^{\mathrm{tSZ}}\approx 4.9f_\nu$, with a larger power spectrum for the `cb' prescription. With our fiducial model this corresponds to $\Delta C_\ell^{\mathrm{tSZ}}/C_\ell^{\mathrm{tSZ}}\approx \Sigma m_\nu/(2.9\,\mathrm{eV})$. This is a very good approximation to the ratios observed in Figure~\ref{fig:pres1} on small angular scales (high multipoles) where neutrino perturbations are negligible compared to baryon and CDM perturbations.  On large angular scales (low multipoles), $P_\mathrm{m}$  also picks up some contributions from neutrino perturbations. Thus the difference between the SZ spectrum computed with or without the `cb' prescription is smaller at low multipole.

With this we conclude that an SZ power spectrum analysis without the `cb' prescription would yield a constraint on $\sigma_8$ biased high by $\approx0.25\%$ if the total neutrino mass was fixed to $\Sigma m_\nu = 0.06$ eV. Thus the `cb' prescription does not appear to be crucial for small neutrino masses, in line with expectations. Note, however, that the total neutrino mass is not strongly constrained by SZ clusters and power spectrum when they are not combined with other data sets. Hence, the maximum likelihood analysis can sample relatively large neutrino masses for which the `cb' prescription becomes important. For instance, our analysis of the Planck SZ cluster counts alone has a best-fitting value at $\Sigma m_\nu = 4.35\,\mathrm{eV}$. For such a large neutrino mass, the predicted number of clusters differs by about 25\% with or without the `cb' prescription. Then, for consistency, it is important to take it into account.

\section{Effect of massive neutrinos on SZ observables}
\label{s:pheno}
The matter power spectrum is damped by massive neutrino free-streaming \citep[e.g.,][]{Eisenstein:1997jh,Hu:1997vi,Lesgourgues:2006nd}. The free-streaming scale is inversely proportional to the total neutrino mass and reaches a maximum at radiation-to-matter equality. Wavelengths larger than the maximum free-streaming scale are not affected by the damping, while wavelengths smaller than the current free-streaming scale are maximally affected. This being said, the global impact of massive neutrinos on cosmological observables is subtle to describe, because a variation of $\Sigma m_\nu$  induces variations of other parameters. These variations and their observable effects depend on which quantities one chooses to keep fixed while varying $\Sigma m_\nu$ (or $\omega_\nu \simeq \Sigma m_\nu/93.14$~eV). We pick three examples:

\begin{enumerate}
\item[\textbf{Case A}:]
We keep the quantities $(A_\mathrm{s}, n_\mathrm{s}, \omega_\mathrm{b}, \Omega_\mathrm{m}, h)$ fixed (and therefore, also $\Omega_\Lambda$ and $\omega_\mathrm{m}$),
while adjusting the CDM density to $\omega_\mathrm{c}=\omega_\mathrm{m}-\omega_\mathrm{b}-\omega_\nu$. This choice is the most conventional one when discussing the effect of neutrino free-streaming on the matter power spectrum. Indeed, it fixes the amplitude of the large-scale power spectrum, and singles out the small-scale step-like suppression induced by massive neutrinos. Note that the redshift of radiation-to-matter equality is not constant in that case. Indeed, for realistic neutrino masses, neutrinos still count as radiation at equality, and $z_\mathrm{eq} = (\omega_\mathrm{b}+\omega_\mathrm{c})/\omega_\mathrm{r}$, where  $\omega_\mathrm{r}$ accounts for the density of photons plus relativistic neutrinos. In this case, when $\Sigma m_\nu$ increases, $\omega_\mathrm{c}$ decreases, and so does $z_\mathrm{eq}$. It is with this choice that the linear total matter power spectrum is suppressed roughly by $(1-8f_\nu)$ in the small-scale limit \citep{Eisenstein:1997jh,Lesgourgues:2006nd}, while the linear `cb' power spectrum is suppressed by $(1-6f_\nu)$ as explained in \cite{Vagnozzi:2018pwo}. Moreover, the actual suppression is not perfectly linear in $f_\nu$ and also depends on the mass splitting. Finally, the quantity $(\sigma_8)^2$, which is of particular interest for the rest of the discussion,  is sensitive to scales on which the neutrino free-streaming effect is not maximal. Using \verb|class|, we find that for $f_\nu = \mathcal{O}(10^{-2})$ and three degenerate massive neutrinos,

$(\sigma_8^\mathrm{m})^2 \propto (1-7.3 f_\nu)\,  \quad \mathrm{while} \quad (\sigma_8^\mathrm{cb})^2 \propto (1-5.5 f_\nu)$~. 

\noindent Meanwhile, the fractional density of baryons and CDM scales simply like $\Omega_\mathrm{cb} \propto (1-f_\nu)$.
\item[\textbf{Case B}:]
We fix the quantities that are best measured by CMB experiments, i.e., the redshift of equality $z_\mathrm{eq}$ (and thus $\omega_\mathrm{c}$) and the angular scale of the sound horizon at decoupling $\theta_\mathrm{s}$. Concretely, one can fix ($A_\mathrm{s}$, $n_\mathrm{s}$, $\omega_\mathrm{b}$, $\omega_\mathrm{c}$, $\theta_\mathrm{s}$), and increase $\Sigma m_\nu$ while decreasing $h$ in such a way to keep a fixed $\theta_\mathrm{s}$. 

\begin{figure}
\begin{centering}
\includegraphics[height=6cm]{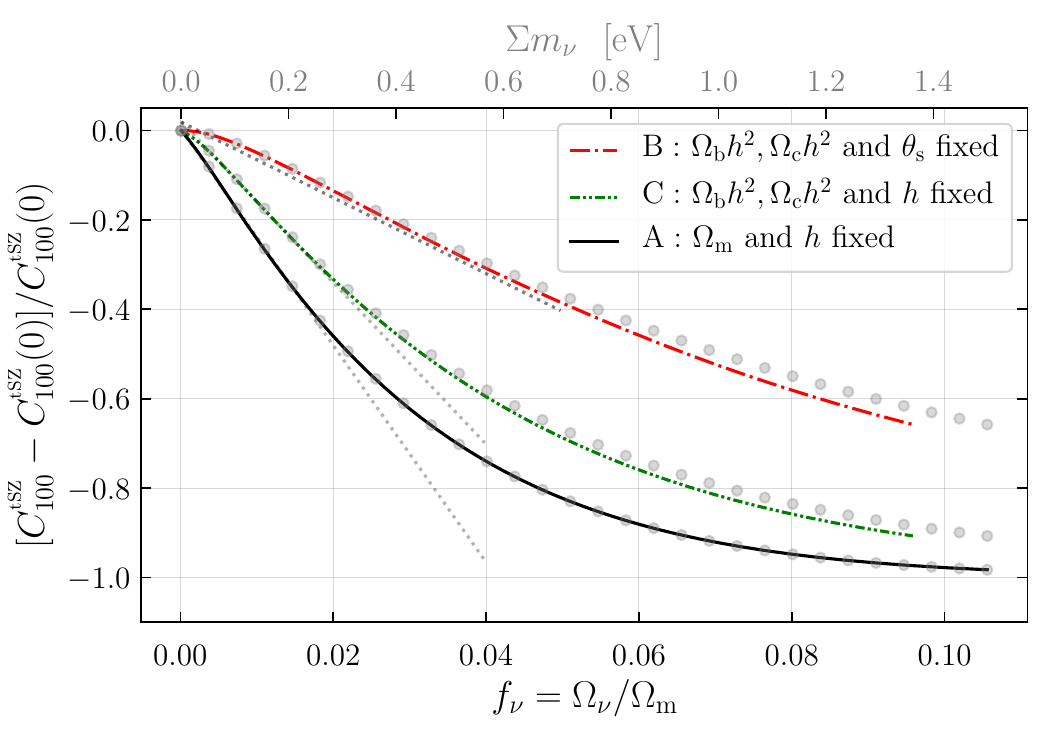}
\par\end{centering}
\caption[]{ Relative variation of the SZ power spectrum amplitude computed at multipole $\ell=100$ with respect to the neutrino fraction $f_\nu$, for different choices of parameterisation referred in the text as case B, C, A. The grey dotted lines are the tangent near the origin with slope $\approx$-8, -18, and -25. The closed grey circles  indicate quantities as a function of $\Sigma m_\nu$.
}
\label{fig:pres2}
\end{figure}
Here, the matter clustering amplitude is suppressed as 

$(\sigma_8^\mathrm{m})^2 \propto (1-6.6 f_\nu) \quad \mathrm{and} \quad (\sigma_8^\mathrm{cb})^2 \propto (1-4.9 f_\nu)$,

\noindent again for $f_\nu = \mathcal{O}(10^{-2})$ and degenerate masses. In this case, maintaining a fixed $\theta_\mathrm{s}$ requires that $h$ varies with the total neutrino mass according to $\Delta h \simeq -0.083 \Delta \left(\Sigma m_\nu/{1~\mathrm{eV}} \right)$, a result consistent with Eq.~(2.1) of \cite{Archidiacono:2016lnv}. For our fiducial parameter values this is equivalent to $h \propto (1-1.6 f_\nu)$.
\item[\textbf{Case C}:]
For illustrative purposes, we mention a third possible choice. One may fix ($A_\mathrm{s}$, $n_\mathrm{s}$, $\omega_\mathrm{b}$, $\omega_\mathrm{c}$, $h$) and increase $ \Sigma m_\nu$: this transformation maintains a fixed redshift of equality, but not a fixed angular scale $\theta_\mathrm{s}$. In this case, the linear matter power spectrum is suppressed by different factors on large and small scales, while $(\sigma_8^\mathrm{m})^2 \propto (1-5.9 f_\nu)$ and $(\sigma_8^\mathrm{cb})^2 \propto (1-4.1 f_\nu)$.
\end{enumerate}
It is then possible to estimate analytically the effect of the total neutrino mass on the SZ power spectrum in each of these cases, and to compare these estimates with the numerical results shown in Figure~\ref{fig:pres2}. The SZ power spectrum scales like 
\begin{equation}
C_\ell^{\mathrm{tSZ}} \propto (\sigma_8^{\mathrm{m}})^{8.1}(\Omega_\mathrm{m})^{3.2}B^{-3.2}h^{-1.7}\,\,\mathrm{for}\,\,\ell\lesssim 10^3\label{eq:clscalm}
\end{equation}
in absence of massive neutrinos \citep[e.g., ][]{Komatsu:2002wc,George:2014oba,Bolliet:2017lha}. So, with the `cb' prescription we now expect
\begin{equation}
C_\ell^{\mathrm{tSZ}} \propto (\sigma_8^{\mathrm{cb}})^{8.1}(\Omega_\mathrm{cb})^{3.2}B^{-3.2}h^{-1.7}\,\,\mathrm{for}\,\,\ell\lesssim 10^3\label{eq:clscal}
\end{equation}
in presence of massive neutrinos. As explained in Appendix \ref{ref:appendix-B}, this scaling and the above discussion leads to  $C_\ell^{\mathrm{tSZ}} \propto (1-\lambda f_\nu)$ with $\lambda\approx 25, 7$ and 17 in case A, B and C respectively, in good agreement with the slopes of Figure~\ref{fig:pres2}.
\newline
\indent In this section we have explored the dependence of the tSZ power spectrum on the neutrino mass (or the neutrino fraction $f_\nu$). To show that the scaling in Eq. \ref{eq:clscal} is valid, we compared the scaling of the tSZ power spectrum obtained numerically when the neutrino mass varies with the scaling obtained semi-analytically, assuming relation \ref{eq:clscal} is correct, and found good agreement. 
\newline
\indent One should be cautious when using a scaling such as $C_\ell^{\mathrm{tSZ}} \propto (1- f_\nu)^\alpha$ derived analytically or from simulations in order to draw conclusions on which value of the neutrino mass can reconcile the relatively low amplitude of tSZ power spectrum (or abundance of clusters) with the amplitude of the primordial power spectrum deduced from CMB: as we highlight here, the value of $\alpha$ depends on the choice of parameterisation.

\begin{figure}
\begin{centering}
\includegraphics[height=5.5cm]{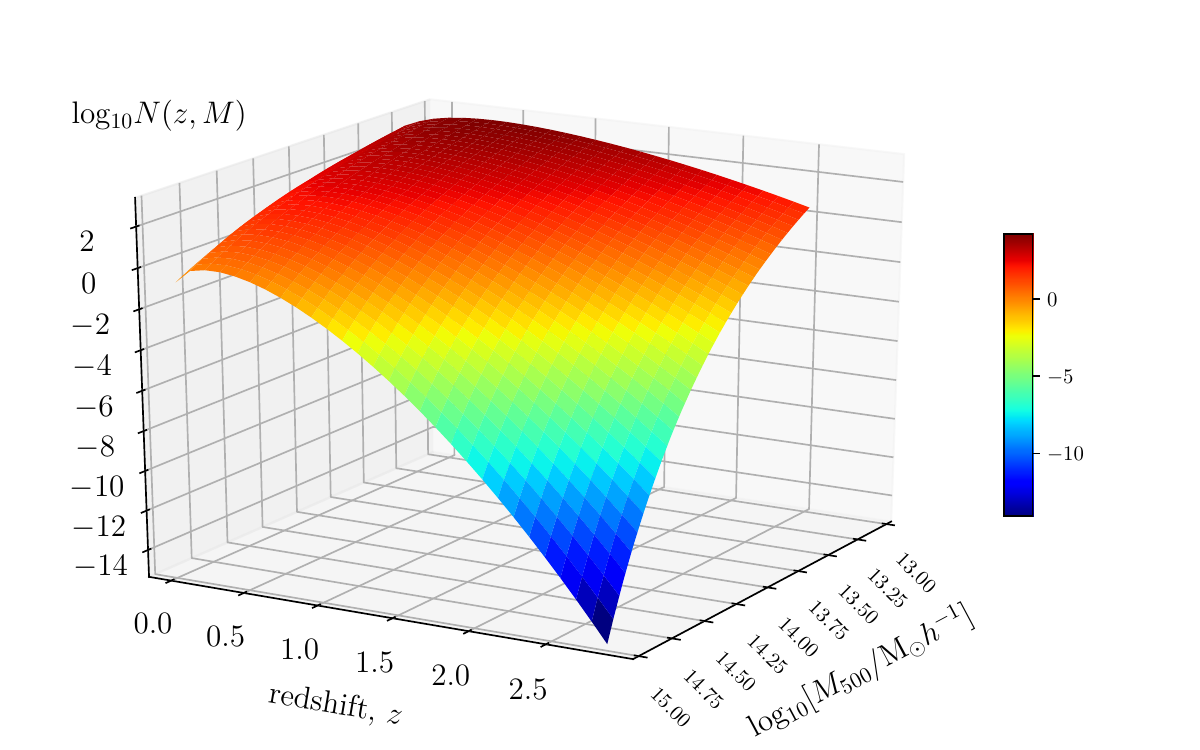}
\includegraphics[height=5.5cm]{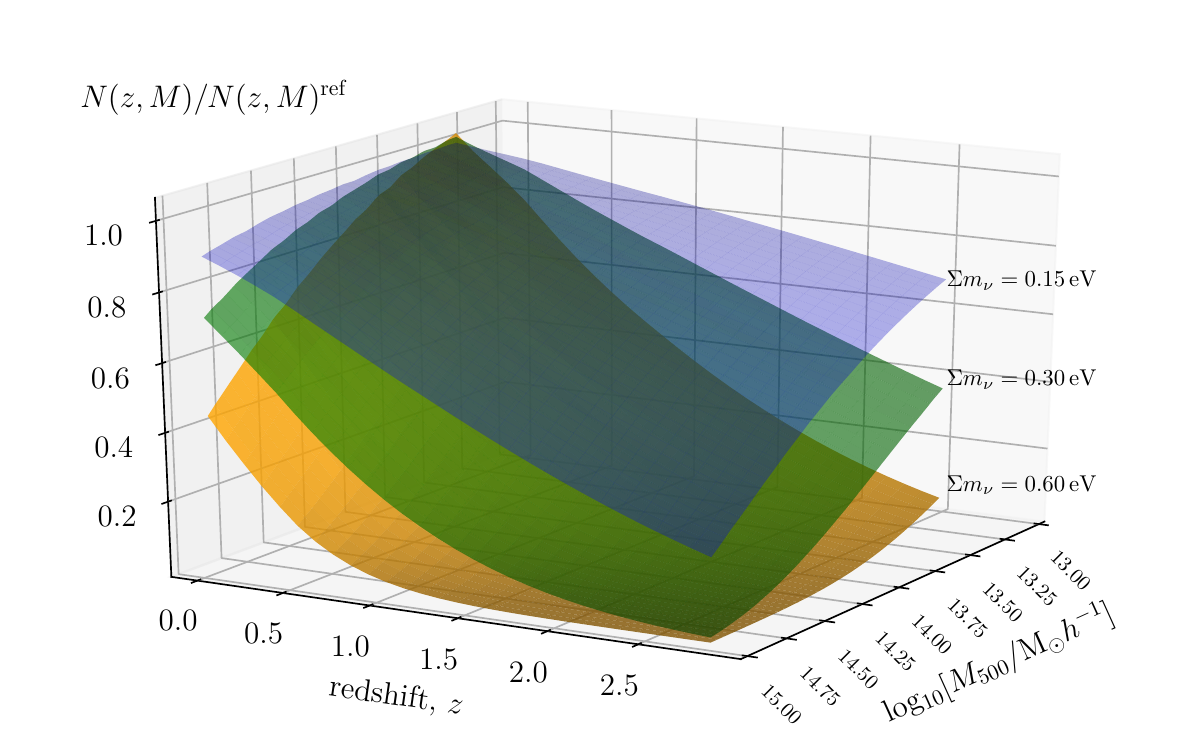}
\par\end{centering}
\caption[]{(Top panel) Cluster counts $N(z,M)$ for  $\Sigma m_\nu=0.06$ eV. (Bottom panel) Variation of $N(z,M)$ with respect to the total neutrino masses. We plot the ratio of $N(z,M)$ computed with three different neutrino masses over the previous one, with the choice of parametrization called `case B' in the text (i.e., fixed $A_\mathrm{s}$, $n_\mathrm{s}$, $\omega_\mathrm{b}$, $\omega_\mathrm{c}$, and $\theta_\mathrm{s}$).}
\label{fig:pheno-3-cc}
\end{figure}

The fact that the scaling of the amplitude of the SZ power spectrum with the total neutrino mass depends on the chosen parameter basis  also applies to cluster counts. For illustrative purposes, we show the effect of the total neutrino mass on $N(z,M)$ for `case B' in figure \ref{fig:pheno-3-cc}. The suppression of cluster abundance is larger at large cluster masses and high redshift because this regime is determined by the exponential tail of the HMF \citep[e.g.,][]{Lukic:2007fc}. 
Although this is not used in this article, for completeness we give the scaling of the total cluster counts for fixed neutrino masses. At fixed total neutrino mass, the scaling of cluster counts with cosmological parameters depends on the survey completeness (as well as the mass and redshift ranges). With \verb|class_sz| we find that the number of clusters scales as
\begin{equation}
N_\mathrm{tot} \propto \sigma_8^{9.8}\Omega_\mathrm{m}^{2.9}B^{-3.2}h^{-0.5}
\end{equation}
where $N_\mathrm{tot}$ is the sum of the clusters in all the redshift and signal-to-noise bins of the Planck analysis. We note that this scaling depends on the survey selection function and leave the detailed analysis for a forthcoming article.

\section{Current constraints}
\label{sec:results}
We study the constraints that can be derived from the Planck SZ data on the $\Lambda$CDM model with a fixed $\Sigma m_\nu =0.06\,\mathrm{eV}$ (Sections \ref{ss:TSZonly}-\ref{ss:SZcomp-b}), on the $\nu\Lambda$CDM model with a varying total neutrino mass (Sections \ref{ss:szcmb}-\ref{ss:szcmbbao}), and finally on the $\nu w$CDM model when additionally varying the dark energy equation of state parameter $w$ (Section \ref{ss:wcdm}). We sample the parameter space with Monte Carlo Markov Chains (MCMCs) using \verb|MontePython|\footnote{\texttt{MontePython} is available at \url{https://github.com/brinckmann/montepython\_public} and the SZ likelihood codes used here are publicly available in the repository.} \citep{2013JCAP...02..001A,Brinckmann:2018cvx}. For the computation of the SZ power spectrum, cluster counts, and CMB spectra we use \verb|class_sz| \citep{Bolliet:2017lha} including the changes described above, i.e., the `cb' prescription (see Section \ref{s:cb}).

\paragraph*{Parameter combinations.}
\label{ss:pc}
The scaling of the SZ power spectrum given in Eq.~\eqref{eq:clscal} motivates the parameter combination
$\sigma_8^{\mathrm{cb}} (\Omega_{\mathrm{cb}} /B)^{0.4}h^{-0.2}$
for the characterisation of the SZ power spectrum amplitude, analogous to the parameter $
F\equiv \sigma_8 (\Omega_{\mathrm{m}} /B)^{0.4}h^{-0.2} $.
used in \cite{Bolliet:2017lha}. (Note that here and hereafter $\sigma_8$ means $\sigma_8^{\mathrm{m}}$.)

In the MCMC analysis of the Planck $y$-map power spectrum data, this is the parameter combination that minimises the relative uncertainty: if we write $F=\sigma_8 (\Omega_{\mathrm{m}} /B)^{\lambda}h^{\gamma}$, we find that $\sigma(F) / F$ is minimal for $\lambda=0.4$ and $\gamma=-0.2$.   

For the Planck SZ cluster counts analyses, we find that the `best' parameter combinations with respect to this criterion are 
$\sigma_8^\mathrm{cb} (\Omega_\mathrm{cb} /B)^{0.35}$ and $\sigma_8 (\Omega_\mathrm{m} /B)^{0.35}$
where we note that the Hubble parameter does not appear. This is consistent with \cite{1993MNRAS.262.1023W} who found on the basis of the Press-Schechter formula that the cluster counts probe mainly the combination  $\sigma_8\Omega_\mathrm{m}^{0.32}$.  When the X-ray mass bias $B$ is fixed or constrained, we also use the parameter combinations $\sigma_8 (\Omega_\mathrm{m}/0.3)^{0.2}$
for the Planck SZ cluster counts analyses \citep[as was used in][for the SPT-SZ analysis]{Bocquet:2018ukq} and $\sigma_8 (\Omega_\mathrm{m}/0.3)^{0.4} h_{70}^{-0.2}$
for the Planck $y$-map power spectrum analyses.  We also discuss constraints on $S_8\equiv \sigma_8 (\Omega_\mathrm{m}/0.3)^{0.5}$ commonly used in large-scale structure analyses.
\\

For an easy outlook on the contents of this section, we summarized our results in Table \ref{ref:table-obs}, with references to the relevant subsections therein.

\setcellgapes{2pt}\makegapedcells
\begin{table}
\begin{centering}
\setcellgapes{2pt}\makegapedcells
\begin{tabular}{c|cccc|cccc|cc}
 Parameter & Min. & Max. &  \tabularnewline
\hline
$10^{9}A_{{\mathrm{s}}}$ & 1.8 & 2.7 \tabularnewline
$n_{{\mathrm{s}}}$ & 0.8 & 1  \tabularnewline
$\tau$ & 0.04 & 0.12  \tabularnewline
$h$ & 0.50 & $>$0.85 \tabularnewline
 $\Omega_{\mathrm{b}}h^{2}$ & 0.0199 & 0.0245 \tabularnewline
 $\Omega_{\mathrm{c}}h^{2}$ & 0.09 & $-$ \tabularnewline
 \hline
 $w$ & $-3$ & $-0.5$ \tabularnewline
 $\Sigma m_\nu\,[\mathrm{eV}]$ & $0$ & $-$ \tabularnewline
\hline
 $B$ & 1 & $-$ \tabularnewline
 $A_{{\mathrm{CIB}}}$ & 0 & 10 \tabularnewline
 $A_{{\mathrm{IR}}}$ & 0 & 10 \tabularnewline
 $a_{{\mathrm{RS}}}$ & -10 & 10 \tabularnewline
 \hline
 $\log_{10} Y_\star$ & -0.3 & -0.08 \tabularnewline
 $\sigma_{Y}$ & 0.02 & 0.12 \tabularnewline
$\alpha$ & 1 & 3 \tabularnewline
\end{tabular}
\par\end{centering}
\caption{Uniform priors used in the maximum likelihood analysis. The notation `$-$' means no bound. The nuisance parameter $a_\mathrm{RS}$ is related to $A_\mathrm{RS}$ of  \protect\cite{Bolliet:2017lha}  via $A_\mathrm{RS}=\exp(-a_\mathrm{RS})$.
}\label{tab:Uniform-priors-imposed}
\end{table}

\begin{table}
\begin{centering}
\begin{tabular}{lc}
 & mean $\pm$ $\sigma$\tabularnewline
\hline 
$\log_{10} Y_\star$ & $-0.19\pm0.02$ \tabularnewline
 $\sigma_Y$ & $0.075\pm0.01$ \tabularnewline
$\alpha$& $1.79\pm0.08$ \tabularnewline

\hline 
\end{tabular}
\par\end{centering}
 \caption{Gaussian priors on the cluster counts likelihood parameters, the same as in \protect\cite{Ade:2015fva}.}\label{tab:addpriors}
\end{table}

\begin{table*}
	\centering{}%
	\begin{tabular}{|c|c||c|c|}
		\hline 
		\multirow{2}{*}{\textbf{Section 5.2}} & tSZ-Y & \multirow{2}{*}{\textbf{Section 5.3}} & tSZ-N\tabularnewline
		& $\Lambda$CDM &  & $\Lambda$CDM\tabularnewline
		\hline 
		$\sigma_{8}\left(\Omega_{\mathrm{m}}/B\right)^{0.4}h^{-0.2}$ & $0.455\pm0.013$ & $\sigma_{8}\left(\Omega_{\mathrm{m}}/B\right)^{0.35}$ & $0.460\pm0.007$\tabularnewline
		\hline 
		\multicolumn{1}{c|}{} & tSZ-Y &  & tSZ-N\tabularnewline
		\multicolumn{1}{c|}{} & $\nu\Lambda$CDM &  & $\nu\Lambda$CDM\tabularnewline
		\hline 
		$\sigma_{8}\left(\Omega_{\mathrm{m}}/B\right)^{0.4}h^{-0.2}$ & prior driven & $\sigma_{8}\left(\Omega_{\mathrm{m}}/B\right)^{0.35}$ & prior driven\tabularnewline
		\hline 
		$\Sigma m_{\nu}$ & prior driven & $\Sigma m_{\nu}$ & prior driven\tabularnewline
		\hline 
		\multicolumn{1}{c}{} & \multicolumn{1}{c}{} & \multicolumn{1}{c}{} & \multicolumn{1}{c}{}\tabularnewline
		\hline
		\multirow{2}{*}{\textbf{Section 5.4}} & tSZ-Y + $B=B_{\mathrm{HSE}}\left(\pm10\%\right)$ & \multirow{2}{*}{\textbf{Section 5.4}} & tSZ-N + $B=B_{\mathrm{HSE}}\left(\pm10\%\right)$\tabularnewline
		& $\Lambda$CDM &  & $\Lambda$CDM\tabularnewline
		\hline 
		$\sigma_{8}\left(\Omega_{\mathrm{m}}/0.3\right)^{0.5}$ & $0.752\pm0.034$ & $\sigma_{8}\left(\Omega_{\mathrm{m}}/0.3\right)^{0.5}$ & $0.764\pm0.029$\tabularnewline
		\hline 
		$\sigma_{8}\left(\Omega_{\mathrm{m}}/B\right)^{0.4}h^{-0.2}$ & $0.756\pm0.033$ & $\sigma_{8}\left(\Omega_{\mathrm{m}}/0.3\right)^{0.2}$ & $0.763\pm0.025$\tabularnewline
		\hline 
		\multicolumn{1}{c}{} & \multicolumn{1}{c}{} & \multicolumn{1}{c}{} & \multicolumn{1}{c}{}\tabularnewline
		\hline
		\textbf{Section 5.5} & tSZ-Y + CMB & \textbf{Section 5.5} & tSZ-N + CMB\tabularnewline
		\hline 
		$\sigma_{8}\left(\Omega_{\mathrm{m}}/B\right)^{0.4}h^{-0.2}$ & $0.469_{-0.003}^{+0.004}$ & $\sigma_{8}\left(\Omega_{\mathrm{m}}/B\right)^{0.35}$ & $0.464\pm0.006$\tabularnewline
		\hline 
		$\left(1-b\right)$ & $0.64_{-0.04}^{+0.03}$ & $\left(1-b\right)$ & $0.62_{-0.04}^{+0.03}$\tabularnewline
		\hline 
		$\Sigma m_{\nu}$ (95\%CL) & $<0.32\,\mathrm{eV}$ & $\Sigma m_{\nu}$ (95\%CL) & $<0.24\,\mathrm{eV}$\tabularnewline
		\hline 
		\multicolumn{1}{c|}{} & tSZ-Y + CMB + $B=B_{\mathrm{HSE}}\left(\pm10\%\right)$ &  & tSZ-N + CMB + $B=B_{\mathrm{HSE}}\left(\pm10\%\right)$\tabularnewline
		\hline 
		$\left(1-b\right)$ & $0.69\pm0.03$ & $\left(1-b\right)$ & $0.68_{-0.04}^{+0.03}$\tabularnewline
		\hline 
		$\Sigma m_{\nu}$ (95\%CL) & $<0.39\,\mathrm{eV}$ & $\Sigma m_{\nu}$ (95\%CL) & $<0.37\,\mathrm{eV}$\tabularnewline
		\hline 
		\multicolumn{1}{c}{} & \multicolumn{1}{c}{} & \multicolumn{1}{c}{} & \multicolumn{1}{c}{}\tabularnewline
		\hline
		\textbf{Section 5.6, 5.7} & tSZ-Y + CMB + BAO & \textbf{Section 5.6, 5.7} & tSZ-N + CMB + BAO\tabularnewline
		\hline 
		$\Sigma m_{\nu}$ (95\%CL) & $<0.127\,\mathrm{eV}$ & $\Sigma m_{\nu}$ (95\%CL) & $<0.129\,\mathrm{eV}$\tabularnewline
		\hline 
		\multicolumn{1}{c|}{} & ... + $B=B_{\mathrm{HSE}}\left(\pm10\%\right)$ &  & ... + $B=B_{\mathrm{HSE}}\left(\pm10\%\right)$\tabularnewline
		\hline 
		$\Sigma m_{\nu}$ (95\%CL) & $<0.149\,\mathrm{eV}$ & $\Sigma m_{\nu}$ (95\%CL) & $<0.164\,\mathrm{eV}$\tabularnewline
		\hline 
		\multicolumn{1}{c|}{} & ... + in $w$CDM  &  & ... + in $w$CDM\tabularnewline
		\hline 
		$\Sigma m_{\nu}$ (95\%CL) & $<0.28\,\mathrm{eV}$ & $\Sigma m_{\nu}$ (95\%CL) & $<0.32\,\mathrm{eV}$\tabularnewline
		\hline 
		$w$ & $-1.07_{-0.06}^{+0.09}$ & $w$ & $-1.09_{-0.07}^{+0.10}$\tabularnewline
		\hline 
	\end{tabular}\caption{Summary of Observational Constraints. Constraints from our analysis of the Planck 2015 y-map are under the acronym tSZ-Y and those from our analysis of Planck 2015 cluster counts are labeled tSZ-N. CMB refers to Planck 2015 primary CMB TT,TE,EE data and likelihood. BAO refers to constraints obtained from the 6dFGS, the SDSS DR7 MGS, and the BOSS DR12 LOWZ \& CMASS galaxy samples. See section \ref{ref:ss-data} for details on the data sets and likelihood codes. The three dots `...'  in the lower part of the table mean the data combination that directly precede from above. The notation $B=B_{\mathrm{HSE}}\left(\pm10\%\right)$ refers to the Gaussian prior for the bias centred on $B_{\mathrm{HSE}}=1.25$ with a 10\% relative width (i.e., $b\approx0.8$).} \label{ref:table-obs}
\end{table*}

\subsection{Data and likelihoods}\label{ref:ss-data}
We use the Planck 2015 $y$-map power spectrum \citep{Aghanim:2015eva}  and Planck 2015 cosmological sample of the SZ cluster catalogue \citep{Ade:2015fva}.  To avoid complications related to the covariance between cluster counts and SZ power spectrum, we do not study joint constraints from the Planck $y$-map power spectrum and Planck SZ cluster counts, but only use them separately. We use uniform priors on input parameters reported in Table \ref{tab:Uniform-priors-imposed}. 
\vspace{5mm}

For the Planck $y$-map power spectrum likelihood we follow \cite{Bolliet:2017lha}, with the same treatment of foreground residuals.  In particular, we fit the Planck 2015 $y$-map power spectrum in  eighteen multipole bins between $\ell_{_\mathrm{min}}=10$ and $\ell_{_\mathrm{max}}=959.5$, including non-Gaussian errors. We sample the three foreground residual amplitudes\footnote{`CIB' refers to the cosmic infrared background, `RS' and `IR' to radio and infra-red  point sources.} ($A_\mathrm{CIB}$, $A_\mathrm{IR}$, $a_\mathrm{RS}$) in addition to the cosmological parameters and X-ray mass bias.

For the Planck SZ cluster counts likelihood, we use the same  
model and likelihood  as the \cite{Ade:2015fva} baseline analysis, i.e., with fixed $\beta=0.66$ and the Gaussian priors on the scaling relations parameters $\alpha$, $\log_{10} Y_\star$ and $\sigma_Y$ reported in Table \ref{tab:addpriors} as well as a detection threshold $\xi_\mathrm{cut}=6$, using information on both signal-to-noise and redshift dimensions. We ported the \verb|CosmoMC| \citep{Lewis:2002ah} likelihood code \verb|szcounts.f90| into  \verb|class_sz|  and \verb|MontePython|. We refer to \cite{Ade:2015fva} and \cite{Zubeldia:2019brr} for a detailed description of the cluster counts likelihood and definitions of $\alpha$, $\log_{10}Y_\star$, $\beta$, $\sigma_Y$.

We also study combinations of the Planck SZ probes with the following datasets:
\begin{itemize}
	\item \textbf{Planck 2015 primary CMB} consisting of Planck 2015 high-$\ell$ temperature and polarisation (TT,TE,EE) plus low-$\ell$ tempature~\citep{Aghanim:2015xee} and a Gaussian prior on the reionisation optical depth\footnote{We never include a low-$\ell$ polarisation likelihood directly and we do not use Planck CMB lensing power spectra.} intended to mimic the constraint from the Planck 2016 low-$\ell$ polarisation likelihood \citep[`SimLow',][]{Aghanim:2016yuo} as in \cite{Calabrese:2017ypx}, i.e. $\tau = 0.06 \pm 0.01$.\footnote{This value agrees with the Planck 2018 results~\citep{Aghanim:2018eyx}, $\tau=0.054\pm 0.08$ (68\% CL).}
	\item \textbf{BAO} consisting of data from the 6dFGS~\citep{Beutler:2011hx}, the SDSS DR7 MGS~\citep{Ross:2014qpa} and the BOSS DR12 LOWZ \& CMASS galaxy samples~\citep{Alam:2016hwk}.
\end{itemize}

The Planck high-$\ell$ TT,TE,EE likelihood has three `nuisance' parameters related to the SZ effect \citep{Ade:2013kta,Aghanim:2015xee}: $\xi^\mathrm{tSZ\times CIB}$ to account for the correlation between SZ and CIB foregrounds; $A^\mathrm{kSZ}$ for the amplitude of the kinetic SZ power spectrum; and $A^\mathrm{tSZ}$ for the amplitude of the thermal SZ power spectrum template. The template for the  thermal SZ power spectrum used in the Planck CMB likelihood is the model of \cite{doi:10.1111/j.1365-2966.2012.21059.x} (with pressure profile evolution parameter $\epsilon=0.5$). Motivated by a SPT measurement \citep{2012ApJ...755...70R}, the Planck CMB likelihood has a conservative Gaussian prior on the linear combination  
\begin{equation}
A^{\mathrm{kSZ}}+1.6A^{\mathrm{tSZ}} = (9.5\pm3)\mu\mathrm{K}^2 \label{eq:gptsz}
\end{equation}
\citep{Ade:2015xua} as well as a uniform prior on $\xi^\mathrm{tSZ\times CIB}$ between 0 and 1 \citep[justified by the SPT constraint of][]{George:2014oba}. In this work, we use the same settings for the SZ templates and nuisance parameters as in the Planck CMB likelihood.\footnote{Although, as emphasised in \cite{Ade:2015xua}, the Gaussian prior of Eq. \eqref{eq:gptsz} and the uniform prior on $\xi^\mathrm{tSZ\times CIB}$ are sufficient to eliminate the sensitivity of the cosmological parameters on the SZ modelling, we note that since the Planck $y$-map power spectrum constrains $A^\mathrm{tSZ}$, the Gaussian prior of Eq. \eqref{eq:gptsz} could be further augmented with a Planck $y$-map motivated prior on $A^\mathrm{tSZ}$.}

\begin{figure}
\begin{centering}
\includegraphics[height=8cm]{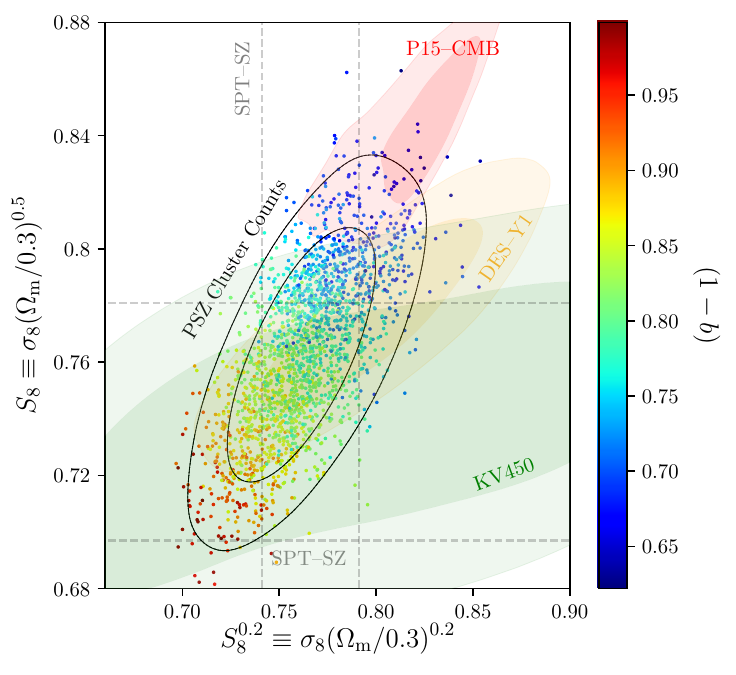}
\par\end{centering}
\caption[]{Marginalised 2d joint posterior probability
 distributions with 68\% CL and 95\% CL contours from the Planck cluster counts analysis with a Gaussian prior on the X-ray mass bias $B=1.250\pm0.125$ (corresponding to $1-b=0.80\pm0.08$) in $\Lambda$CDM (black contours). The red Planck contours `P15-CMB' are from our chains in $\nu\Lambda$CDM. The orange `DES-Y1' contours are from \protect\citep{Abbott:2017wau} in $\nu\Lambda$CDM `3x2pt'. The green  contours `KV-450' for KiDS+VIKING are from \citep{Hildebrandt:2018yau} in $\Lambda$CDM. The dashed lines show the 1-$\sigma$ intervals from the SPT-SZ cluster counts analysis \protect\citep[`SPTcl',][]{Bocquet:2018ukq}.

}
\label{fig:compspt}
\end{figure}

\subsection{Planck $y$-map power spectrum alone}\label{ss:TSZonly}
In the flat $\Lambda$CDM model with $\Sigma m_\nu = 0.06\,\mathrm{eV}$,  we measure the parameter combination $\sigma_8 (\Omega_{\mathrm{m}} /B)^{0.4}h^{-0.2}$ to a $\approx 3$\% precision:
\begin{equation}
\sigma_8 (\Omega_{\mathrm{m}} /B)^{0.4}h^{-0.2}=0.455\pm0.013\,\, (68\%\,\mathrm{CL}).\nonumber\label{eq:Fco}
\end{equation}
Without the `cb' prescription, we find $\sigma_8 (\Omega_{\mathrm{m}} /B)^{0.4}h^{-0.2}=0.456\pm0.013$ (68\% CL), i.e., a negligible $\approx 0.2\%$ upward shift, as anticipated in Section~\ref{s:cb}. We note that our constraint on this parameter combination in $\Lambda$CDM, without the `cb' prescription, is 1\% lower than the one quoted \cite{Bolliet:2017lha}. This small difference is due to the new version of \verb|class_sz| which uses an original \verb|class| routine for the computation of $\sigma(R)$ instead of the method based on Chebyshev polynomial interpolation of the previous version. 

In $\nu\Lambda$CDM, the total neutrino mass is not constrained by the Planck $y$-map power spectrum data alone. In our analysis, the 95\% CL interval for $\Sigma m_\nu$ is determined by the upper bound of the $H_0$ prior. At large total neutrino mass there is an anti-correlation between $\Sigma m_\nu$ and $\sigma_8^{\mathrm{cb}} (\Omega_{\mathrm{cb}} /B)^{0.4}h^{-0.2}$, so the $\nu\Lambda$CDM constraint on this parameter combination is also prior driven. Nevertheless,  for $\Sigma m_\nu< 0.1\,\mathrm{eV}$, the 68\% and 95\% CL intervals are independent of $\Sigma m_\nu$ with $\sigma_8^{\mathrm{cb}} (\Omega_{\mathrm{cb}} /B)^{0.4}h^{-0.2}=0.456\pm 0.013$  (68\% CL).

\vspace{-3mm}
\subsection{Planck SZ cluster counts alone}\label{ss:SZcounts-only}
In the flat $\Lambda$CDM model with $\Sigma m_\nu = 0.06\,\mathrm{eV}$, we measure the parameter combination $\sigma_8 (\Omega_\mathrm{m} /B)^{0.35}$ to a $\approx$1.5\% precision:
\begin{equation}
\sigma_8 (\Omega_\mathrm{m} /B)^{0.35}=0.460\pm0.007\,\,  (68\%\,\mathrm{CL}).\nonumber\label{eq:Fstarco}
\end{equation}
If we do not adopt the `cb' prescription, we obtain a value of $\sigma_8 (\Omega_\mathrm{m} /B)^{0.35}$ that is $\approx 0.2\%$  larger, as also found for the Planck $y$-map power spectrum analysis.

In $\nu\Lambda$CDM, the total neutrino mass is unconstrained by the SZ cluster counts alone. Like in the Planck $y$-map power spectrum analysis, we find that the 95\% CL interval for $\Sigma m_\nu$ is determined by the upper bound of the $H_0$ prior. Thus, the $\nu\Lambda$CDM constraint on $\sigma_8^\mathrm{cb} (\Omega_\mathrm{cb} /B)^{0.35}$ is also prior driven. Nevertheless, for $\Sigma m_\nu< 0.1\,\mathrm{eV}$, the 68\% and 95\% CL intervals of $\sigma_8^\mathrm{cb} (\Omega_\mathrm{cb} /B)^{0.35}$ are almost independent of $\Sigma m_\nu$ with $\sigma_8^\mathrm{cb} (\Omega_\mathrm{cb} /B)^{0.35}=0.461\pm 0.007$~(68\%~CL).  

\begin{figure}
\begin{centering}
\includegraphics[height=5cm]{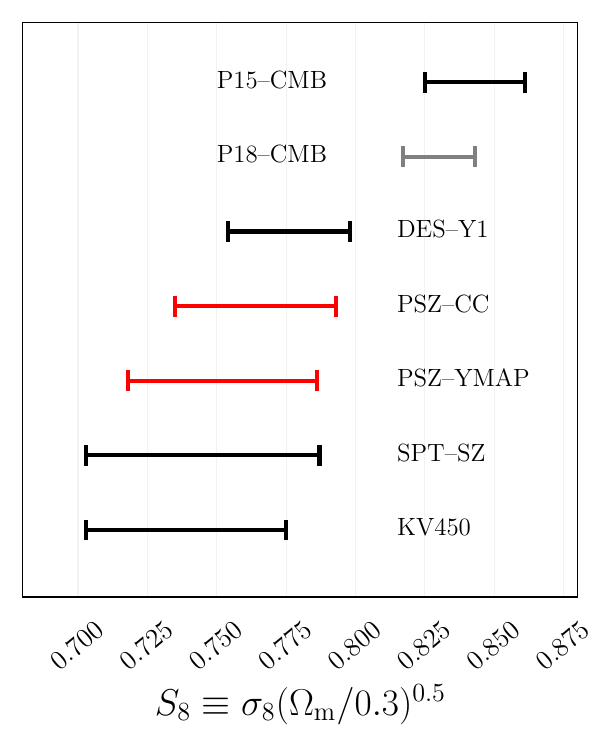}\includegraphics[height=5cm]{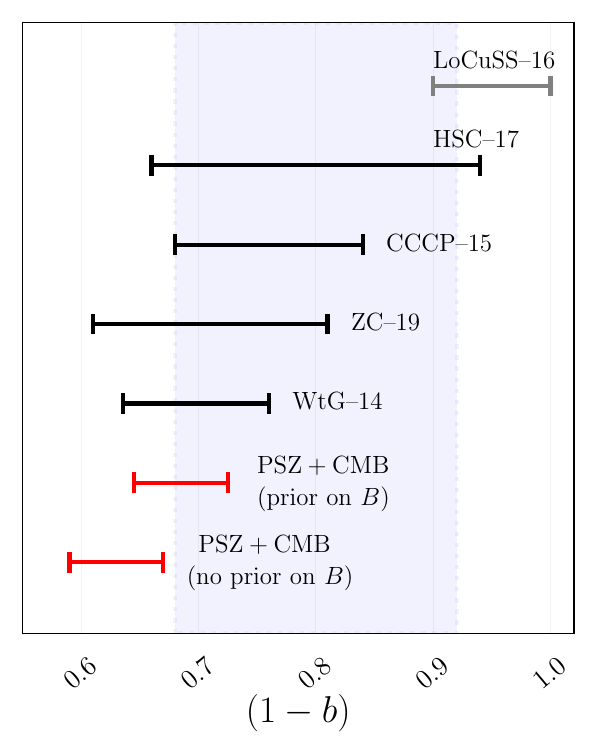}
\par\end{centering}
\caption[]{(Left panel) Marginalised 1-$\sigma$ constraints on $S_8$ from several datasets. The grey constraints `P18-CMB' refers to the latest $\Lambda$CDM measurement from \protect\citep{Aghanim:2018eyx}. See caption of Figure~\ref{fig:compspt} for the others. (Right panel)  Marginalised 1-$\sigma$ constraints on $(1-b)\equiv B^{-1}$. The blue band is the simulation result of \protect\cite{Shi:2015fua} (see their Figure 5, `NG all'). Also reported are Weighing the Giants  \citep[WtG,][]{vonderLinden:2014haa},  Canadian Cluster Comparison Project  \citep[CCCP,][]{Hoekstra:2015gda}, Subaru Hyper Suprime-Cam \citep[HSC,][]{HSC},  Local Cluster Substructure Survey \citep[LoCuSS,][]{Smith:2015qhs}, and the updated Planck CMB weak lensing constraint \citep[ZC,][]{Zubeldia:2019brr}. The `PSZ' values are the median central values of the Planck SZ cluster counts and Planck SZ power spectrum analyses. }
\label{fig:comps8}
\end{figure}

\begin{figure}
\begin{centering}
\includegraphics[height=8cm]{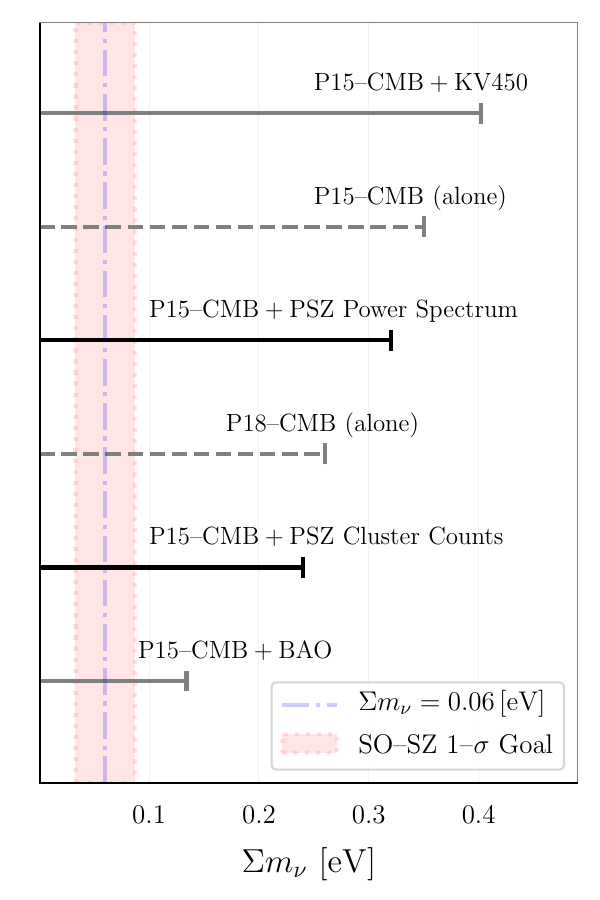}
\par\end{centering}
\caption[]{Constraints on the total neutrino mass from several datasets. All but the Planck 2018 constraint are the results of our MCMC analyses. The P18-CMB constraint is $\Sigma m_\nu <0.26\,\mathrm{eV}$ (95\% CL,  TT,TE,EE+lowE) is from \cite{Aghanim:2018eyx} for . The P15-CMB+KV450 constraint is $\Sigma m_\nu <0.40\,\mathrm{eV}$ (95\% CL), for Planck 2015 CMB combined with KiDS+VIKING-450 \citep{Hildebrandt:2018yau}. The P15-CMB+BAO is $\Sigma m_\nu <0.13\,\mathrm{eV}$ (95\% CL, Section \ref{ss:szcmbbao}). The P15-CMB and SZ constraints are the first quoted in section \ref{ss:szcmb}, with no prior on the mass bias for the SZ constraints. For further comparison, the pink shaded area corresponds to the 1-$\sigma$ goal from the \cite{Ade:2018sbj}, i.e., $\sigma(\Sigma m_\nu)=27$ meV.}
\label{fig:comp-mnu}
\end{figure}

\subsection{Planck SZ with prior on the X-ray mass bias}\label{ss:SZcomp-b}
Motivated by the discussion of Section~\ref{ref:ssb} we set a Gaussian prior on the X-ray mass bias centred on $B_\mathrm{HSE}=1/(1-b_\mathrm{HSE})$ with a 10\% relative width, i.e., $B =  1.250 \pm 0.125$. With this prior, our analysis of Planck SZ cluster counts in $\Lambda$CDM with $\Sigma m_\nu = 0.06\,\mathrm{eV}$ yields
\begin{eqnarray}
\sigma_8 (\Omega_\mathrm{m}/0.3)^{0.2}&=&0.763\pm 0.025\,\, (68\%\,\mathrm{CL}),\nonumber\\
 \sigma_8 (\Omega_\mathrm{m}/0.3)^{0.5} &=&0.764\pm 0.029\,\, (68\%\,\mathrm{CL}).\nonumber
\end{eqnarray}
Our analysis of Planck $y$-map power spectrum in $\Lambda$CDM with $\Sigma m_\nu = 0.06\,\mathrm{eV}$ yields
\begin{eqnarray}
\sigma_8 (\Omega_\mathrm{m}/0.3)^{0.4} h_{70}^{-0.2}&=&0.756\pm 0.033\,\, (68\%\,\mathrm{CL}),\nonumber\\
\sigma_8 (\Omega_\mathrm{m}/0.3)^{0.5} &=&0.752\pm 0.034\,\, (68\%\,\mathrm{CL}).\nonumber
\end{eqnarray}
As illustrated on Figure~\ref{fig:compspt} and on the left panel of Figure~\ref{fig:comps8}, the Planck SZ constraints are fully consistent with the results of SPT-SZ \citep{Bocquet:2018ukq}, DES-Y1 \citep{Abbott:2017wau} and KiDS+VIKING-450 \citep{Hildebrandt:2018yau}, and in mild tension with Planck 2015 primary CMB \citep[see][for a discussion of the moderate tension between DES-Y1 and Planck]{Handley:2019wlz}. There are three straightforward ways to alleviate the small mismatch between Planck SZ and Planck 2015 primary CMB: a larger X-ray mass bias, more-massive neutrinos or a less-negative $w$. Our analyses presented hereafter suggest that the larger X-ray mass bias is preferred by the Planck data.

\subsection{Planck SZ with Planck 2015 primary CMB}\label{ss:szcmb}
In $\nu\Lambda$CDM, for Planck 2015 primary CMB alone we find $\Sigma m_\nu< 0.35\,\mathrm{eV}\,\,(\mathrm{95\%\,CL})$. When we do not use a prior constraint on the X-ray mass bias, the addition of Planck SZ improves the 2-$\sigma$ limit slightly. For Planck 2015 primary CMB combined with Planck SZ cluster counts we obtain
\[
\begin{array}{cclc}
\sigma_8 (\Omega_\mathrm{m} /B)^{0.35}&=&0.464\pm0.006\,\,&(68\%\,\mathrm{CL}),\\
(1-b)&=& 0.62^{+0.03}_{-0.04}\,\,&(68\%\,\mathrm{CL}),\\
\Sigma m_\nu&<&0.24 \,\mathrm{eV}\,\,&(\mathrm{95\%\,CL,\,\,\mathrm{no\,\,prior\,\,on}\,\,}B).
\end{array}
\]
With  Planck $y$-map power spectrum we obtain
\[
\begin{array}{cclc}
 \sigma_8 (\Omega_{\mathrm{m}} /B)^{0.4}h^{-0.2}&=&0.469^{+0.004}_{-0.003}\,&(68\%\,\mathrm{CL}),\\
(1-b)&=& 0.64^{+0.03}_{-0.04}\,&(68\%\,\mathrm{CL}),\\
\Sigma m_\nu&<&0.32 \,\mathrm{eV}\,&(\mathrm{95\%\,CL,\,\mathrm{no\,prior\,on}\,}B).
\end{array}
\]
These results show a remarkable consistency between the Planck SZ cluster counts and $y$-map power spectrum data: constraints on the X-ray mass bias are consistent with each other at the half-$\sigma$ level. 

\begin{figure}
\begin{centering}
\includegraphics[height=8.5cm]{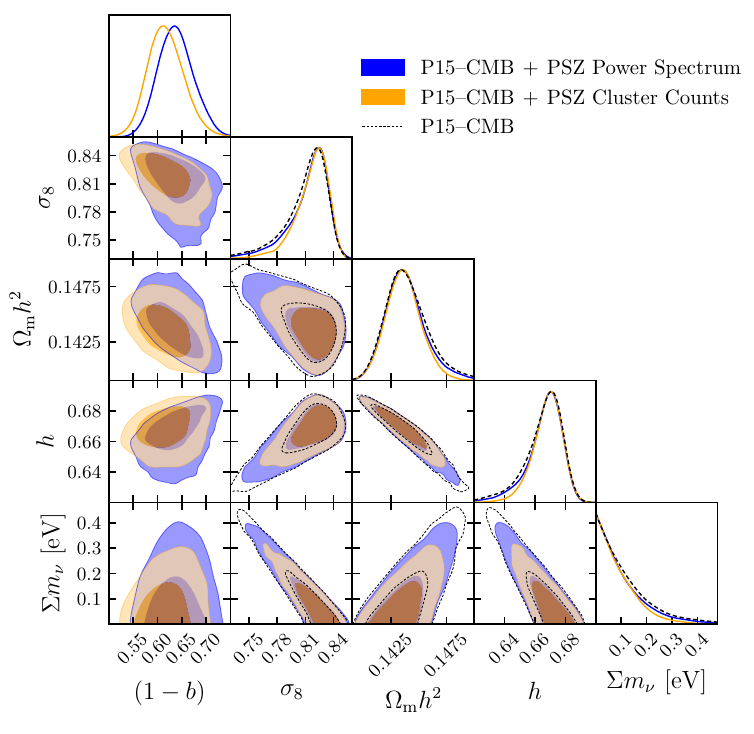}
\par\end{centering}
\caption[]{Marginalised (1d and 2d) joint posterior probability
 distributions with 68\% CL and 95\% CL contours (for a sub-set of parameters) obtained from the Planck $y$-map power spectrum (blue) and Planck SZ cluster counts (orange) combined with Planck 2015 primary CMB  in $\nu\Lambda$CDM. Contours from  Planck 2015 primary CMB alone are the dashed lines. No prior information is assumed for the mass bias, other than a flat prior for  $B=1/(1-b)$.
}
\label{fig:tSZ+CMBcomp}
\end{figure}

Contours for both analyses, as well as for Planck 2015 primary CMB alone, are shown on Figure~\ref{fig:tSZ+CMBcomp} for a  sub-set of parameters. We see that the resulting constraints for the X-ray mass bias are not consistent with the HSE  value $(1-b)\approx 0.8$.

One could argue that leaving the X-ray mass bias unconstrained is not relevant as this is equivalent to saying that the cluster masses are unknown. We therefore performed analyses where we used the Gaussian prior of Section~\ref{ss:SZcomp-b} on the bias, i.e., $B =  1.250 \pm 0.125$.  The resulting constraints on the parameter combinations $\sigma_8 (\Omega_\mathrm{m} /B)^{0.35}$ and $\sigma_8 (\Omega_{\mathrm{m}} /B)^{0.4}h^{-0.2}$ are the same as without the X-ray mass bias prior. However, the 2-$\sigma$ limit of the total neutrino mass becomes weaker. More precisely, with the X-ray mass bias prior, for Planck 2015 primary CMB combined with Planck SZ cluster counts we obtain
\[
\begin{array}{cclc}
(1-b)&=& 0.68^{+0.03}_{-0.04}\,\,&(68\%\,\mathrm{CL}),\\
\Sigma m_\nu&<&0.37 \,\mathrm{eV}\,\,&(\mathrm{95\%\,CL,\,\,\mathrm{prior\,\,on}\,\,}B).
\end{array}
\]
With Planck $y$-map power spectrum we obtain
\[
\begin{array}{cclc}
(1-b)&=& 0.69\pm 0.03\,\,&(68\%\,\mathrm{CL}),\\
\Sigma m_\nu&<&0.39 \,\mathrm{eV}\,\,&(\mathrm{95\%\,CL,\,\,\mathrm{prior\,\,on}\,\,}B).
\end{array}
\]
The 1d posterior probability distribution of the total neutrino mass extends towards larger values when we adopt a Gaussian prior on the X-ray mass bias centred on $B_\mathrm{HSE}$, because, in this case, the Planck SZ data is consistent with a matter clustering significantly lower than the one favoured by Planck 2015 primary CMB (see left panel of Figure~\ref{fig:comps8}).  Hence, a larger neutrino mass helps to bring the primary CMB inferred $\sigma_8$ towards lower values, more consistent with the SZ inferred  $\sigma_8$. Nevertheless, with our bias prior, the combination of SZ with Planck 2015 primary CMB does not single-out a particular value of the total neutrino mass. On the contrary, we find that the data combination prefers pushing the X-ray mass bias to the upper end of the Gaussian prior, rather than increasing the neutrino mass. Unless there is stringent observational evidence for a relatively small mass bias (which would correspond to imposing a tight Gaussian prior around $(1-b)=0.8$), our results suggest that the neutrino mass does not help reducing the inconsistency between tSZ and CMB probes\footnote{As reported in \cite{McCarthy:2017csu} we confirmed in one of our analyses that when the mass bias is \textit{fixed} to $(1-b)=0.8$, Planck SZ and Planck CMB are consistent with one another with a total neutrino mass $\Sigma m_\nu \simeq 0.4$ eV. This is an interesting result, however, current data does not allow to fix the mass bias parameter to an arbitrary precision, making the conclusion drawn from this case somewhat premature.}. We summarised our results on the X-ray mass bias from Planck SZ plus Planck 2015 primary CMB in the right panel of Figure~\ref{fig:comps8} along with several other constraints from weak lensing cluster mass calibrations, to show the current level of uncertainty on this parameter. We note that CMB weak lensing mass calibration methods with forthcoming experiments such as CMB-S4 are expected to yield a determination of the X-ray mass bias at the percent level \citep[see, e.g.,][]{Louis:2016gvv}.

\subsection{Planck SZ with Planck 2015 primary CMB and BAO}\label{ss:szcmbbao}
In $\nu\Lambda$CDM, for Planck 2015 primary CMB combined with BAO we find $\Sigma m_\nu< 0.134\,\mathrm{eV}\,\,(\mathrm{95\%\,CL})$.   This is consistent with the results of \cite{Alam:2016hwk}, i.e., $\Sigma m_\nu< 0.16\,\mathrm{eV}\,\,(\mathrm{95\%\,CL})$, although slightly stronger because we use a prior on $\tau$ rather than the `lowP' likelihood.\footnote{Low-$\ell$ polarisation TE,EE,BB likelihood.} When we do not use a prior constraint on the X-ray mass bias, the addition of Planck SZ improves the 2-$\sigma$ limit only by a few percent. More precisely, with Planck SZ cluster counts we obtain  
\begin{equation}
\Sigma m_\nu<0.129\,\,\mathrm{eV}\,\,(\mathrm{95\%\,CL,\,\,\mathrm{no\,\,prior\,\,on}\,\,}B),\nonumber
\end{equation}
 and with Planck $y$-map power spectrum we obtain 
 \begin{equation}
 \Sigma m_\nu<0.127\,\,\mathrm{eV}\,\,(\mathrm{95\%\,CL,\,\,\mathrm{no\,\,prior\,\,on}\,\,}B).\nonumber
 \end{equation}
 Like in the previous section, with a Gaussian prior on the X-ray mass bias, the addition of SZ to Planck 2015 primary CMB and BAO  slightly  degrades the 2-$\sigma$ limit on the total neutrino mass.  Indeed, when we use the Gaussian prior on the X-ray mass bias of Section~\ref{ss:SZcomp-b}, we obtain $$\Sigma m_\nu< 0.164\,\mathrm{eV}\,\,(\mathrm{95\%\,CL,\,\,\mathrm{prior\,\,on}\,\,}B)$$ with Planck SZ cluster counts and $$\Sigma m_\nu< 0.149\,\mathrm{eV}\,\,(\mathrm{95\%\,CL,\,\,\mathrm{prior\,\,on}\,\,}B)$$ with Planck $y$-map power spectrum.  Again, we do not find that the marginalised posterior probability distribution of the total neutrino mass peaks at a particular value in any of these analyses. 
Regarding the X-ray mass bias, we obtain the same constraints as in the  previous section at the half-$\sigma$ level, i.e.,  using Planck SZ plus Planck 2015 primary CMB.
\newline

Constraints on the total neutrino mass in $\Lambda$CDM for the combination of Planck SZ data with CMB are shown against other relevant measurements in Figure \ref{fig:comp-mnu}. It shows that adding SZ to CMB results in a slightly better constraint on the neutrino mass, competitive with current analysis of galaxy cosmic shear combined with CMB. But the improvement on the neutrino mass bound is not as stringent as the combination of CMB and BAO. This is because the low redshift distance measurements provided by BAO break the degeneracy between the effects of massive neutrino and the Hubble parameter on the primary CMB temperature anisotropy power spectrum. While the anti-correlation between $h$ and $\Sigma m_\nu$ is almost unchanged when adding SZ data to Planck 2015 CMB (see bottom right corner of Figure \ref{fig:tSZ+CMBcomp}). For this reason, the addition of SZ data to CMB+BAO only marginally improves the constraint on $\Sigma m_\nu$ compared to CMB+BAO (from 134 meV to 129 meV with cluster counts and from 134 meV to 127 meV with SZ power spectrum data, all at 95\% CL), as obtained in subsection \ref{ss:szcmbbao}.

\subsection{Massive neutrinos and dark energy}\label{ss:wcdm}
Before presenting the joint constraints from Planck SZ with CMB and BAO data in $\nu w$CDM, we update the analysis of \cite{Bolliet:2017lha}, i.e., the measurement of the equation of state of dark energy  from the Planck $y$-map power spectrum by using only minimal information from primary CMB consisting of a normalisation prior on $A_\mathrm{s}e^{-2\tau}$, as well as the prior on the optical depth $\tau$, complemented by a measurement  of the Hubble constant. 
\paragraph*{Planck $y$-map with minimal CMB information.}
Once $H_0$ is fixed, a correlation between $w$ and the angular size of the sound horizon at decoupling arises. Hence, the prior on  $\theta_\mathrm{s}$ used in \cite{Bolliet:2017lha} is crucial for  the measurement of $w$. But since the Planck 2015 primary CMB constraint on $\theta_\mathrm{s}$ is model-independent, i.e., it is essentially the same in models with dark energy or massive neutrinos \citep[see Table 5 of][]{Aghanim:2018eyx}, we can use this information to improve the  constraint. We consider the latest measurement of the Hubble constant from \cite{Riess:2019cxk} and use a Gaussian prior on the X-ray mass bias, $B=1.410\pm0.141$ corresponding to $(1-b)\approx 0.7$, consistent with the results of our joint analyses of Planck 2015 primary CMB with Planck SZ with constrained bias and also with \cite{Zubeldia:2019brr}. In Table \ref{tab:priorw}, we report the priors and external measurements for this analysis.

 In $\Lambda$CDM with $\Sigma m_\nu = 0.06$ eV, we obtain the constraint
$w = -1.22_{-0.08}^{+0.10}\,\,(68\%\,\mathrm{CL})$ from the analysis of the Planck $y$-map power spectrum.  We note that the mean value depends on our choices of X-ray mass bias and Hubble constant. For instance, a smaller bias $B$, or a smaller $H_0$, would drive $w$ towards less-negative values. The same method can be applied to find a constraint on the total neutrino mass. In $\nu\Lambda$CDM ($w=-1$) with the same settings we find 
$
\Sigma m_\nu < 0.17\,\mathrm{eV}\,\,(95\%\,\mathrm{CL}),
$ almost the same as with full CMB information. However, in $\nu w$CDM we note that this method does not allow for a constraint on either of these parameters because of the degeneracy between  $w$   and $\Sigma m_\nu$: the effects of a more-negative $w$ can be compensated by more-massive neutrinos.

\paragraph*{Planck SZ plus CMB and BAO.}
Due to an additional degeneracy between  $w$   and $\Sigma m_\nu$, when we do the same analyses as in the previous section (combining Planck SZ cluster counts or Planck $y$-map power spectrum with Planck 2015 primary CMB, BAO data and with the Gaussian prior on the X-ray mass bias around $B_\mathrm{HSE}$) in $\nu w$CDM,  the 2-$\sigma$ limit on the total neutrino mass becomes significantly weaker than in $\nu\Lambda$CDM (see Section~\ref{ss:szcmbbao}). More precisely, with Planck $y$-map  power spectrum we obtain 
 \[
\begin{array}{cclc}
\Sigma m_\nu&<& 0.28\,\mathrm{eV}\,\,&(95\%\,\mathrm{CL}),\\
w&=&-1.07^{+0.09}_{-0.06}\,\,&(\mathrm{68\%\,CL)},
\end{array}
\]
and with Planck SZ cluster counts we obtain 
 \[
\begin{array}{cclc}
\Sigma m_\nu&<& 0.32\,\mathrm{eV}\,\,&(95\%\,\mathrm{CL}),\\
w&=&-1.09^{+0.10}_{-0.07}\,\,&(\mathrm{68\%\,CL)},
\end{array}
\]
in agreement with the standard $\Lambda$CDM model.
For the X-ray mass bias, we obtain the same constraints as in Section~\ref{ss:szcmb} at the half-$\sigma$ level, i.e., using Planck SZ plus Planck 2015 primary CMB with Gaussian prior on $B$.
\newline

\indent Overall, our results on the neutrino mass, dark energy equation of state, and cosmological parameters are consistent with the standard $\Lambda$CDM model: we find that neither dark energy or massive neutrino help in bridging the gap between the relatively low $S_8$ from SZ data and the relatively high value derived from CMB (see left panel of Figure \ref{fig:comps8} for a summary of the constraints on $S_8$).  The preferred solution to this mild discrepancy seems to be a slightly larger X-ray mass bias than the one expected from hydrodynamical simulations (see right panel of Figure \ref{fig:comps8}).
 
The results of this section are summarised in Table \ref{ref:table-obs} at the start of Section~\ref{sec:results}.

\begin{table}
\begin{centering}
\begin{tabular}{lc}
 & mean $\pm$ $\sigma$\tabularnewline
\hline 
$10^9 A_\mathrm{s} e^{-2\tau}$& $1.878\pm0.014$ \tabularnewline
$\tau$& $0.06\pm0.01$ \tabularnewline
$h$& $0.7403\pm0.0142$ \tabularnewline
$100 \theta_\mathrm{s}$& $1.04093\pm0.00030$ \tabularnewline
$B$&$1.410\pm0.141$\tabularnewline
\hline 
\end{tabular}
\par\end{centering}
\caption{Gaussian priors for the analyses of Section \ref{ss:wcdm}. The normalisation prior is the one derived in \protect\cite{Bolliet:2017lha}, the $\tau$ prior is from \protect\cite{Calabrese:2017ypx}, the $h$ prior is from \protect\cite{Riess:2019cxk}, the $\theta_\mathrm{s}$ prior is the same as in \protect\cite{Zubeldia:2019brr}. The X-ray mass bias prior is motivated by the results of Section~\ref{ss:szcmb}, and those of \protect\cite{Zubeldia:2019brr}.}\label{tab:priorw}
\end{table}

\section{Forecasts}\label{s:f}
To assess the constraining power of cosmic variance limited SZ power spectrum measurements we use the MCMC method, expected to be more reliable and accurate than the Fisher matrix method \citep[e.g., ][for a comparison of both methods]{Perotto:2006rj} especially when the posterior likelihood probability distribution is non-Gaussian, as is the case here (see contours of Figure~\ref{fig:tszcvlsp}). We describe the settings of our analyses in Section~\ref{subs:sets} and present our results in Section~\ref{ss:cvl-ps}. The results of this section are summarised in Table \ref{ref:table-forecast}.

\begin{table*}
	\centering{}%
	\begin{tabular}{|c|c|c|}
		\hline 
		\textbf{Section 6.2} & 10\% prior on mass bias & 1\% prior on mass bias\tabularnewline
		\cline{1-1} 
		CMB-S4 + tSZ-Y ($\ell_{\mathrm{max}}=10^{4}$) & masking ($M<5\times10^{14}\mathrm{M}_{\varodot}/h$) & masking ($M<5\times10^{14}\mathrm{M}_{\varodot}/h$)\tabularnewline
		\hline 
		$\sigma(\Sigma m_{\nu})$ & $31$ meV & $29$ meV\tabularnewline
		\hline 
		\hline 
		... + DESI-BAO & no masking & masking ($M<5\times10^{14}\mathrm{M}_{\varodot}/h$)\tabularnewline
		\hline 
		$\sigma(\Sigma m_{\nu})$ & $26$ meV & $24$ meV\tabularnewline
		\hline 
	\end{tabular}\caption{Summary of Forecasted Constraints. Forecasted uncertainty on the total neutrino mass using our mock tSZ power spectrum likelihood with $\ell_\mathrm{max}=10^4$ combined with mock CMB-S4 (top row) and combined with mock CMB-S4+DESI-BAO (bottom row). See Section \ref{subs:sets} for details on the data sets and likelihood codes. The difference between left and right on the top row is the relative width of the prior imposed on the mass bias, and on the bottom row it is whether heavy clusters are masked or not. (For the combination CMB-S4+DESI-BAO+tSZ the prior on the bias makes no difference in the forecasted uncertainty.)}\label{ref:table-forecast}
\end{table*}

\subsection{Analyses settings}\label{subs:sets}
To generate realistic mock SZ power spectrum data one should in principle draw the SZ power spectrum data points from the $n$-point probability distribution function of the  SZ power spectrum (where $n$ is the number of multipole bins). \cite{Zhang:2007psa} presented the theoretical framework to compute it. They explained that the  evaluation of the probability distribution function  when $n$ is large requires sophisticated Monte-Carlo integration methods, but that for a large sky coverage at sufficiently large $\ell$  the probability distribution function is well approximated by a multivariate Gaussian distribution. Hence, we bin our mock data from $\ell_\mathrm{min}=200$ to $\ell_\mathrm{max}=10^3$, or $10^4$, with constant logarithmic bin width $\Delta \ln \ell =0.3$ and use the same sky fraction as the Planck $y$-map, i.e., $f_\mathrm{sky}=0.47$. 

To random draw the mock SZ power spectrum data, we use the Cholesky decomposition of the covariance matrix, i.e., $\mathrm{M=LL^t}$ where $\mathrm{L}$ is a lower triangular matrix with positive diagonal elements. So that the mock power spectrum is given by $\hat{C}_\ell=C_\ell^{\mathrm{mock}}+s_\ell$ where $C_\ell^{\mathrm{mock}}$ is the SZ power spectrum calculated for the the fiducial model, and $\mathbf{\mathrm{\mathbf{s}}}=\mathrm{L\mathbf{v}}$ with $\mathrm{\mathbf{v}}$ a vector of $n$ random numbers drawn from the Normal distribution.\footnote{The non-Gaussian statistics of the tSZ power spectrum induces multipole-to-multipole correlations which can bias the centre of the confidence contours when using only one realisation of the mock data to study forecasts. A way to quantify this bias would be to study the results of many realisations. Nevertheless, here we are interested in the width of the contours, which we do not expect to be biased when we use just one realisation.}

\begin{figure}
\begin{centering}
\includegraphics[height=6.cm]{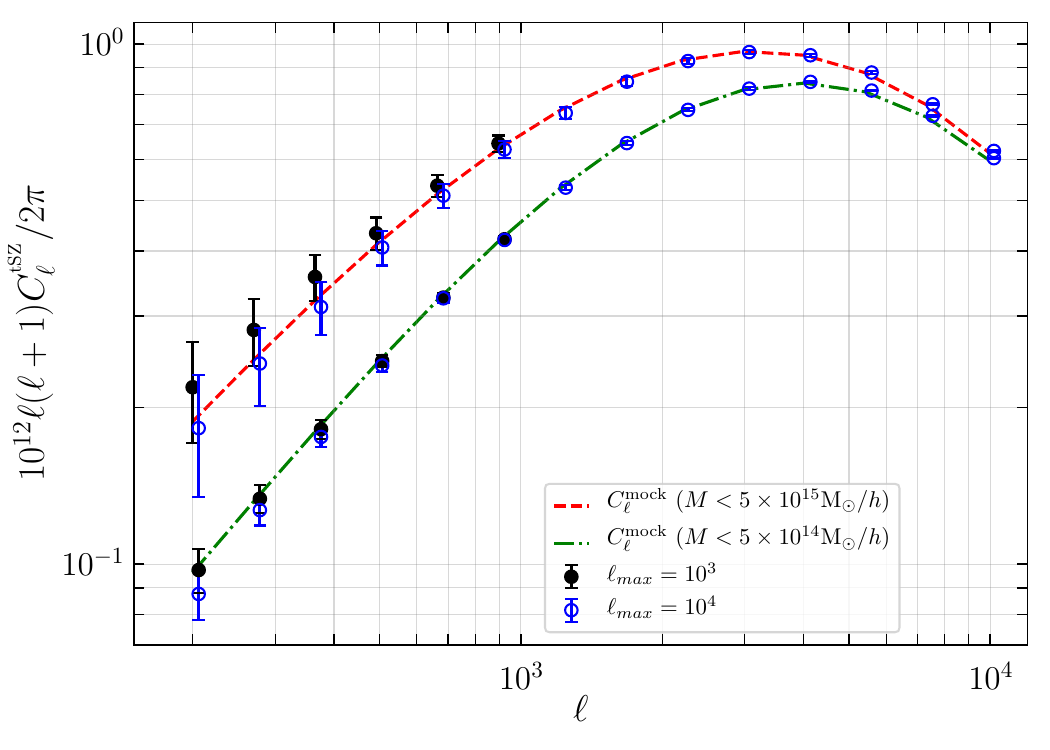}
\includegraphics[height=6cm]{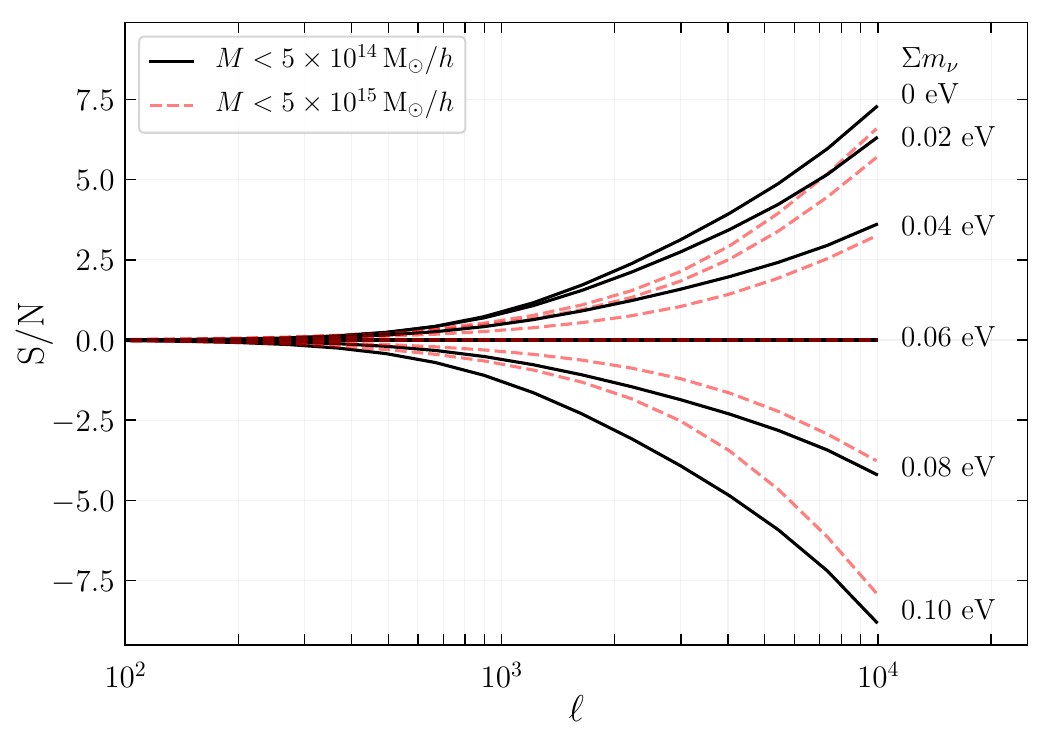}
\includegraphics[height=6cm]{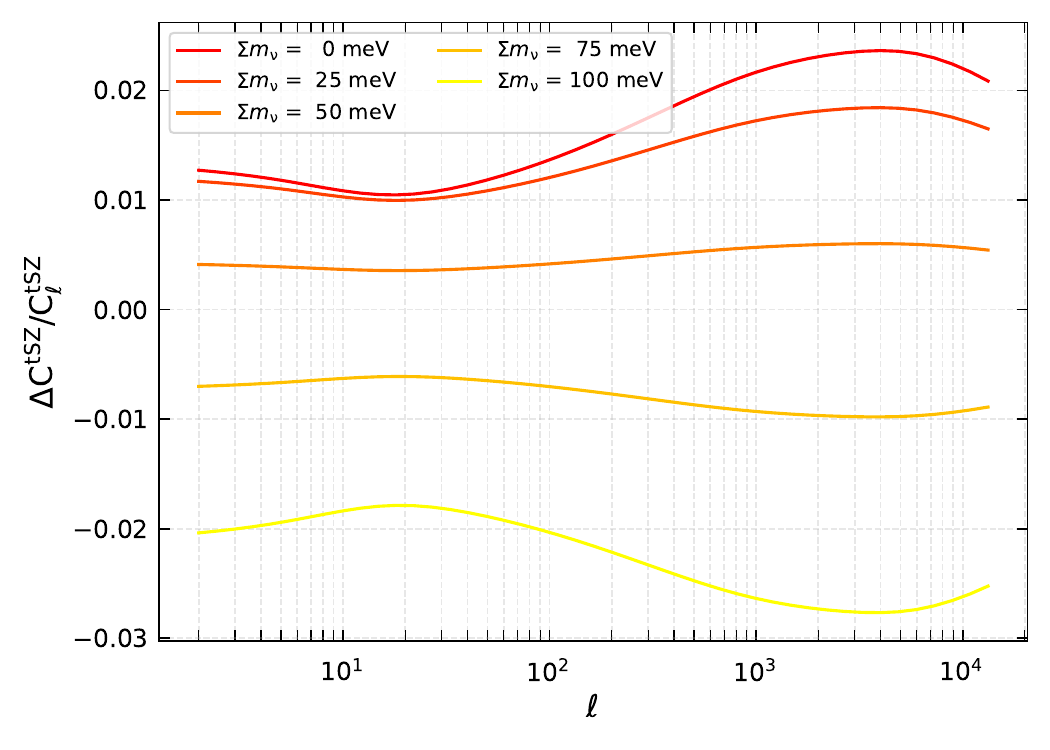}
\par\end{centering}
\caption[]{(Top panel) Mock SZ power spectra with error bars for $\ell_\mathrm{max}=10^3$ and $\ell_\mathrm{max}=10^4$, and the models used to compute them (Planck 2018 cosmology and X-ray mass bias $B=1.41$) for two different  values of $M_\mathrm{max}$. (Middle panel) Variation of signal-to-noise `S/N' of the tSZ power spectrum at each multipole with respect to a fiducial total neutrino mass at $\Sigma m_\nu =0.06$ eV for two different  values of $M_\mathrm{max}$. Here by signal-to-noise we mean the ratio of $\Delta C_\ell=C_\ell(\Sigma m_\nu)-C_\ell(0.06\,\mathrm{eV})$ by the statistical uncertainty, i.e., Gaussian and non-Gaussian sampling variance. Note that we use $f_\mathrm{sky}=0.47$. (Bottom panel) Relative variation of the tSZ power spectrum with the total neutrino mass. The y-axis is $\Delta C_\ell/C_\ell=[C_\ell(\Sigma m_\nu)-C_\ell(0.06\,\mathrm{eV})]/C_\ell(0.06\,\mathrm{eV})$. While varying the neutrino mass we kept the parameters ($A_\mathrm{s}$, $n_\mathrm{s}$, $\omega_\mathrm{b}$, $\omega_\mathrm{c}$, $\theta_\mathrm{s}$) fixed, i.e., `Case B' of section \ref{s:pheno}.}
\label{fig:cov}
\end{figure}

For the fiducial model we use the Planck 2018 cosmological parameters \citep[fifth column of Table 2 of ][]{Aghanim:2018eyx} with a total neutrino mass $\Sigma m_{\nu}=0.06\,\,\mathrm{eV}$ and a reference bias $B=1.41$.  On the top panel of Figure~\ref{fig:cov} we show our mock SZ power spectrum data. In this figure, in particular at large angular scales ($\ell<10^3$), we can see a multipole-to-multipole correlation between the mock SZ data points which is due to the large non-Gaussian contributions to the covariance matrix (trispectrum). It is possible to reduce the trispectrum with respect to the tSZ power spectrum by masking the heaviest clusters \citep[see, e.g.,][ and middle panel of our Figure~\ref{fig:cov}]{Hill:2013baa}. Hence, we also consider a simple masking scenario, for illustration, assuming that all clusters above a certain mass $M_\mathrm{cut}$ have been detected and masked, irrespective of their redshift. For $M_\mathrm{cut}=5\times10^{14}\mathrm{M_\varodot}/h$, this could be achieved by combining information from forthcoming X-ray surveys
 such as eROSITA with the SZ catalogue of clusters of The Simons Observatory. The implementation of this masking scenario is obtained by cutting the mass integral at $M_\mathrm{max}=5\times 10^{14}\mathrm{M}_\odot/h$ (instead of $M_\mathrm{max}=5\times 10^{15}\mathrm{M}_\odot /h$, the upper bound that we normally use in our computations). Assuming the Planck 2018 cosmology, this excludes $\approx950$ clusters over the entire sky.

\cite{Kodwani_2019} argued that using a fixed covariance matrix is generally more consistent than using a covariance that varies at each steps of the MCMC. In the MCMC analysis, we therefore use a fixed covariance matrix, with Gaussian and non-Gaussian (trispectrum) contribution computed with the parameters of the fiducial model. \footnote{In principle, one should also follow this for the Planck $y$-map power spectrum analysis of the previous section. We checked that this difference in methodology does not yield significant changes in the parameter estimation: within a few percent, consistent with \citet[][]{Kodwani_2019}.}

To obtain competitive constraints on the neutrino masses, it is necessary to combine the SZ power spectrum with CMB and potentially BAO data. For the CMB we use a combination of a mock Planck and a mock CMB-S4 likelihood from \cite{Brinckmann:2018owf}, with mock Planck comparable in sensitivity to the final Planck release~\cite{Aghanim:2018eyx} and CMB-S4 specifications from~\cite{Abazajian:2016yjj}. The mock CMB combination is only temperature and polarisation (i.e., no lensing power spectrum) and consists of Planck up to $\ell \leq 50$ with 57\% sky coverage, CMB-S4 at $50 < \ell < 3000$ with 40\% sky coverage plus Planck with 17\% sky coverage (intended to avoid regions of overlapping data). See \cite{Brinckmann:2018owf} for more details on the likelihoods. For BAO we use a mock DESI likelihood as in~\cite{Archidiacono:2016lnv} with redshifts $0.15 < z < 1.85$ known at percent level precision~\citep{Font-Ribera:2013rwa,Allison:2015qca}. We emphasise that in combination with the mock SZ power spectrum we use the `2pt' CMB information, i.e., without the lensing power spectrum (`4pt'). Nevertheless, for comparison, we also quote the `4pt' CMB-S4 constraint.

\subsection{Forecasted constraints}\label{ss:cvl-ps}

For Planck plus CMB-S4, the sensitivity does not enable a determination of $\Sigma m_\nu$ with statistical significance but an upper bound: $\Sigma m_\nu < 0.18\,\mathrm{eV}$ (95\% CL). For Planck plus CMB-S4 with lensing power spectrum we obtain  $\Sigma m_\nu < 0.16\,\mathrm{eV}$ (95\% CL). When we combine Planck plus CMB-S4 with BAO we find a determination of the total neutrino mass with $\sigma(\Sigma m_\nu)^{\mathrm{2pt}}=28\,\mathrm{meV}$. The addition of lensing power spectrum to  CMB and BAO yields $\sigma(\Sigma m_\nu)^{\mathrm{4pt}}=25\,\mathrm{meV}$, consistent with published CMB-S4 forecasted constraints  \citep{Abazajian:2016yjj,Brinckmann:2018owf} as well as the \cite{Ade:2018sbj} goals.

\subsubsection*{CMB-S4+SZ versus CMB-S4+BAO}
For mock SZ power spectrum with Planck plus CMB-S4, we obtain an  improvement on the total-neutrino mass constraints over Planck plus CMB-S4 in the following cases. With mock SZ power spectrum up to $\ell_\mathrm{max}=10^3$, a 1\% prior on the X-ray mass bias and masking, we obtain $\Sigma m_\nu < 0.16\,\mathrm{eV}\,\, (95\%\,\,\mathrm{CL})$. With $\ell_\mathrm{max}=10^4$, we obtain 
$$\sigma(\Sigma m_\nu)=31\,\mathrm{meV}\,\, (\mathrm{\ell_\mathrm{max}=10^4,\,\,masking,\,\,10\%\,\,prior\,\,on\,\,}B),$$
$$\sigma(\Sigma m_\nu)=29\,\mathrm{meV}\,\, (\mathrm{\ell_\mathrm{max}=10^4,\,\,masking,\,\,1\%\,\,prior\,\,on\,\,}B).$$

The effect of a tighter prior on the X-ray mass bias is relatively small: from 10\% to 1\% precision on $B$ the neutrino mass constraint improves by only 2 meV. Indeed a change of the mass bias amounts to a nearly scale invariant rescaling of the SZ power spectrum \citep[see, e.g., Figure~2 of][]{Bolliet:2017lha}, while a change of the total neutrino mass  is not a scale invariant effect, as can be seen on the bottom panel of Figure~\ref{fig:cov}. Hence, with $\ell_\mathrm{max}=10^4$, the X-ray mass bias and total neutrino mass are no longer as degenerate as for $\ell_\mathrm{max}=10^3$. 

\begin{figure}
\begin{centering}
\includegraphics[height=8.2cm]{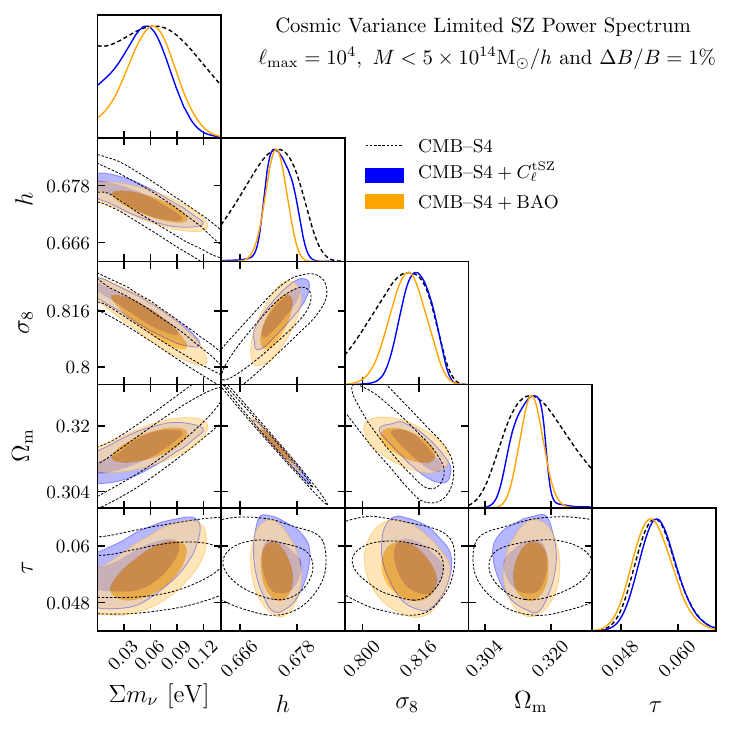}
\par\end{centering}
\caption[]{Marginalised (1d and 2d) joint posterior probability
 distributions with 68\% CL and 95\% CL contours (for a sub-set of parameters) obtained from the mock cosmic variance limited SZ power spectrum combined with CMB-S4 in $\nu\Lambda$CDM (blue).  The X-ray mass bias is constrained with a Gaussian prior $B=1.41\pm 0.014$. Contours for CMB-S4 alone are the dashed lines. Contours for CMB-S4 plus DESI-BAO are in orange. Note that CMB-S4 also includes Planck as described in the last paragraph of Section~\ref{subs:sets}, and does not include lensing power spectrum.}
\label{fig:tszcvlsp}
\end{figure}

Contours for the analysis with 1\% precision on the mass bias, for a sub-set of parameters, are shown on Figure~\ref{fig:tszcvlsp} along with Planck plus CMB-S4 and Planck plus CMB-S4 with BAO contours. Note that the 1\% precision on the X-ray mass bias is motivated by forecasted sensitivity of CMB-S4 experiments \citep[see, e.g.,][]{Louis:2016gvv}. Moreover, note that our forecasted  sensitivity to the optical depth mainly comes from our mock low-$\ell$ TE,EE polarisation data yielding $\sigma(\tau)=0.004$, slightly stronger than the final \cite{Aghanim:2018eyx} constraint but weaker than the CLASS experiment projected sensitivity \citep{Watts:2018etg}.

These results show that masking the heaviest clusters and measuring the SZ power at small scales combined with CMB-S4 `2pt' can yield constraints on the neutrino mass that are nearly as tight as CMB-S4 `2pt' combined with DESI-BAO. Since BAO and SZ do not have the same systemics this means that future SZ power spectrum measurements can potentially become an independent complementary probe of the total neutrino mass. However, we recall that the small-scale SZ power spectrum depends on the details of the ICM pressure profiles, that are not yet fully understood and measured \citep[e.g.,][]{doi:10.1111/j.1365-2966.2012.21059.x,Ruppin:2019deo}. Nevertheless, it appears that combining SZ power spectrum with CMB, in the same way as we did here, can yield tight constraints on ICM properties. With our analyses, we can quantify this at the level of the X-ray mass bias parameter.  Using mock Planck plus CMB-S4 and cosmic variance limited SZ power spectrum with $\ell_\mathrm{max}=10^3$ and the 10\% precision X-ray mass bias prior, we recover the fiducial X-ray mass bias value with $\sigma(B)/B=1.9\%$ without masking. Thus,  we can safely say that the future of the SZ power spectrum as a probe of the ICM properties is  bright.  

\subsubsection*{CMB-S4+BAO+SZ versus CMB-S4+BAO}

For mock SZ power spectrum with CMB and BAO, we obtain an improvement of the total-neutrino mass constraints upon CMB plus BAO when we use $\ell_\mathrm{max}=10^4$. In this case, the analyses with the 1\% and 10\% priors on the X-ray mass bias give the same results, for the reasons explained above. Without masking we obtain
$$\sigma(\Sigma m_\nu)=26\,\mathrm{meV}\,\, (\mathrm{\ell_\mathrm{max}=10^4,\,\,no\,\,masking,\,\,BAO}).$$
With masking we obtain $$\sigma(\Sigma m_\nu)=24\,\mathrm{meV}\,\, (\mathrm{\ell_\mathrm{max}=10^4,\,\,masking,\,\,BAO}),$$ i.e., a $\approx 14\%$ improvement over CMB plus BAO, and competitive with the Planck plus CMB-S4 `4pt' forecast (including the lensing power spectrum). Again, this improvement also relies on the assumption that ICM properties are known well enough so that the SZ power spectrum at $\ell_\mathrm{max}\approx10^4$  can be used as a cosmological probe. 

The results of this section are summarised in Table \ref{ref:table-forecast}.

\section{Discussion and Conclusions}\label{sec:conclusion}
In this work, we have consistently included the effect of massive neutrinos into the theoretical models for the SZ power spectrum and cluster counts, in line with the works of  \cite{Ichiki:2011ue}, \cite{Costanzi:2013bhatex} and \cite{LoVerde:2014rxa}.
\begin{itemize}
\item We updated the scaling of the tSZ power spectrum amplitude and found that for $f_\nu = \omega_\nu/\omega_\mathrm{m} \ll 1$, the amplitude is well described by 
\begin{equation}
C_\ell^{\mathrm{tSZ}} \propto (\sigma_8^{\mathrm{cb}})^{8.1}(\Omega_\mathrm{cb})^{3.2}B^{-3.2}h^{-1.7}\,\,\mathrm{for}\,\,\ell\lesssim 10^3\nonumber
\end{equation}
where $\sigma_8^{\mathrm{cb}}$ and $\Omega_\mathrm{cb} = \Omega_\mathrm{c}+\Omega_\mathrm{b}$ contains only the contribution of CDM and baryons. We note that the scaling with the Hubble parameter depends on the model for the electron pressure profile for the ICM. Here we used the \cite{2013A&A...550A.131P} pressure profile, as implemented in \cite{Bolliet:2017lha}.

\item We identified parameter combinations that minimise the relative uncertainty for both the SZ power spectrum analysis and the cluster counts analysis of the Planck SZ data. Explicitly, these are $\sigma_8^{\mathrm{cb}} (\Omega_{\mathrm{cb}} /B)^{0.4}h^{-0.2}$ for the Planck $y$-map power spectrum and $\sigma_8^{\mathrm{cb}} (\Omega_{\mathrm{cb}} /B)^{0.35}$ for the Planck SZ cluster counts, where $\sigma_8^{\mathrm{cb}}$ indicates that $\sigma_8$ is computed with the power spectrum of CDM and baryons rather than with the total matter power spectrum and $B=(1-b)^{-1}$ is the X-ray mass bias. 

\item For neutrino masses  $\Sigma m_\nu\lesssim 1.5\,\,\mathrm{eV}$ we found  that the parameter combinations with $\sigma_8^{\mathrm{cb}}$ and $\Omega_{\mathrm{cb}}$ are not correlated with the total neutrino mass, and therefore remain well constrained in $\nu\Lambda$CDM with same values as in $\Lambda$CDM. In $\Lambda$CDM, we obtained the tight constraints $\sigma_8^{\mathrm{cb}} (\Omega_{\mathrm{cb}} /B)^{0.4}h^{-0.2}=0.456\pm 0.013$ (68\%CL) from the Planck $y$-map power spectrum alone and $\sigma_8^{\mathrm{cb}} (\Omega_{\mathrm{cb}} /B)^{0.35}=0.461\pm 0.007$ (68\%CL) from the Planck SZ cluster counts alone. 

\item In $\Lambda$CDM, we fixed the X-ray mass bias to $B_\mathrm{HSE}$ with a 10\% precision Gaussian prior, i.e., $B=1.250\pm0.125$, and obtained constraints on $\sigma_8 (\Omega_\mathrm{m}/0.3)^{0.5}$, $\sigma_8 (\Omega_\mathrm{m}/0.3)^{0.2}$, and $\sigma_8 (\Omega_\mathrm{m}/0.3)^{0.4} h_{70}^{-0.2}$ from the Planck cluster counts and Planck $y$-map power spectrum analyses in $\Lambda$CDM  (see Figure~\ref{fig:compspt}). We found that the Planck SZ constraints are consistent with results from SPT-SZ, KiDS+VIKING-450 and DES-Y1 and in mild tension with Planck 2015 primary CMB.

\item In $\nu\Lambda$CDM, when we leave the X-ray mass bias  free and combine the Planck SZ data with the Planck 2015 primary CMB data, we obtained constraints on the X-ray mass bias that are not consistent with the typical departure from hydrostatic equilibrium calculated by hydrodynamical simulations, i.e., $B\simeq 1.25$ or equivalently $(1-b)\simeq0.8$. From Planck 2015 primary CMB with Planck $y$-map power spectrum, we obtained  $(1-b)=0.64^{+0.03}_{-0.04}$ (68\%CL), and with Planck cluster counts we obtained $(1-b)=0.62^{+0.03}_{-0.04}$ (68\%CL) . These values agree within 1-$\sigma$ with other published results \citep[e.g.,][]{Ade:2015fva,Makiya:2018pda,Salvati:2017rsn}.

\item We found that the addition of the SZ data to Planck 2015 primary CMB improves the neutrino mass constraint marginally. Our strongest constraint is with SZ cluster counts, namely $\Sigma m_\nu<0.24 \,\mathrm{eV}\,\,(\mathrm{95\%CL})$. Nevertheless, if we set a Gaussian prior on the bias, the total neutrino mass constraint becomes weaker (see last paragraph of Section~\ref{ss:szcmb}). The same is true when we combine Planck SZ data with CMB and BAO (see Section~\ref{ss:szcmbbao}). 

\item We carried out analyses of the Planck SZ data combined with Planck primary CMB and BAO in $\nu w$CDM (Section~\ref{ss:wcdm}) with Gaussian prior on the X-ray mass bias centred around $B_\mathrm{HSE}=1.25$. Our results suggest that this combination of data prefers to adjust the X-ray mass bias to the higher end of the Gaussian prior on $B$, rather than increasing the total neutrino mass or decreasing $w$, to accommodate the SZ data with primary CMB.\footnote{We note that it is only when the X-ray mass bias is \textit{fixed} to  $B_\mathrm{HSE}=1.25$ that we recover a preference for a total neutrino mass around $\Sigma m_\nu=0.4~$eV, as was obtained by \cite{McCarthy:2017csu}.} 
\item Our results suggest that the apparent tension between Planck SZ data and Planck 2015 primary CMB should be solved by an X-ray mass bias larger than $B_\mathrm{HSE}$, consistent with the one recently obtained from the CMB lensing calibration of the Planck SZ clusters in the analysis of \cite{Zubeldia:2019brr} (see right panel of our   Figure~\ref{fig:comps8} for a summary of the constraints on the X-ray mass bias).  For instance, a larger X-ray mass bias $B$ could be explained by the expected departure from HSE of $\approx 20\%$ \textit{plus} X-ray temperature calibration bias of $\approx 15\%$ for the most massive clusters \citep[see the detailed review by][for a thorough description of current issues related to the cluster mass scale]{Pratt:2019cnf}. Given current uncertainties on the X-ray mass bias (see right panel of Figure~\ref{fig:comps8}), this appears as a better solution rather than extending the standard cosmological model. Nevertheless, adjusting the X-ray mass bias would not explain why other recent results from  SZ clusters and galaxy surveys have also reported a low matter clustering amplitude compared to primary CMB (see left panel of Figure~\ref{fig:comps8}). Hence, extensions to the minimal $\Lambda$CDM cosmology are still worth exploring, especially if they can also explain the difference between  local measurements of the Hubble constant \citep[e.g.,][]{Riess:2019cxk} and those obtained from Planck primary CMB \citep{Aghanim:2018eyx}. Three examples of such extensions are the local void idea \citep[e.g.,][]{Ichiki:2015gia}, interacting dark matter-dark radiation \citep[e.g.,][]{Buen-Abad:2017gxg} and self-interacting neutrinos \citep[e.g.,][]{Kreisch:2019yzn}.
\\

To explore the constraining power of future SZ power spectrum measurements, we used mock cosmic variance limited SZ power spectrum experiments up to $\ell_\mathrm{max}=10^3$ and $\ell_\mathrm{max}=10^4$. The choice $\ell_\mathrm{max}=10^3$ typically reflects the sensitivity of current measurements, while the choice $\ell_\mathrm{max}=10^4$ reflects the targeted sensitivity of next generation CMB experiments \citep[e.g.,][]{Ade:2018sbj,Basu:2019rzm}.

\item We found that to improve upon the neutrino mass constraint from Planck plus CMB-S4 combined with DESI-BAO, it is necessary to  measure the small-scale power spectrum up to $\ell_\mathrm{max}=10^4$ and mask the the heaviest clusters. 
For $\ell_\mathrm{max}=10^4$, we found that the precision on the X-ray mass bias does not affect the total neutrino mass constraint when we combine with Planck plus CMB-S4 and BAO. This is because once the amplitude of primordial curvature perturbations $A_\mathrm{s}$ is fixed, the  effect of neutrino mass  on the SZ power spectrum at small scales differs from the effect at large scales, whereas the effect of the bias is close to a scale-invariant shift of the amplitude. 

\item For the combination of SZ power spectrum with Planck plus CMB-S4 without BAO, we found that to reach a constraint on total neutrino mass competitive with Planck plus CMB-S4 plus DESI-BAO, it will be necessary to mask the heaviest clusters and to probe the SZ power spectrum up to $\ell_\mathrm{max}=10^4$.  In this case, we obtained our tightest forecasted constraint $\sigma(\Sigma m_\nu)=29\,\mathrm{meV}$ when we further assumed a 1\% determination of the X-ray mass bias $B$ (\cite{Louis:2016gvv} have shown that this level of precision can be achieved using CMB-S4 weak lensing masses\footnote{To do so, the authors compare the tSZ flux $Y_{500}$, which gives the X-ray mass (assuming the Y-M relation is calibrated with X-ray data), with the lensing convergence field which gives the lensing mass. Assuming the lensing mass is an unbiased estimator of the true mass, the ratio of both masses yields a 1\% determination of the X-ray mass bias. See \cite{Louis:2016gvv} for details on the forecast and \cite{Melin:2014uaa} for more details on the method.}). This is potentially an important result since SZ power spectrum and BAO do not have the same systematics, making the neutrino mass determination a robust one.

Nevertheless, the SZ power spectrum at small scales is not only determined by cosmological expansion and perturbations but also by the details of the ICM via the pressure profile.  Hence, it may be challenging to use the SZ power spectrum up to $\ell_\mathrm{max}=10^4$  as an independent probe of the total neutrino mass. However, the small-scale SZ power spectrum can instead be used as a powerful probe of the ICM's properties: our mock cosmic variance limited SZ power spectrum experiments easily yield a measurement of the X-ray mass bias at the 2\% precision level when we use information from mock Planck plus CMB-S4 experiments to constrain the cosmological parameters.  In addition to probing the ICM, other promising avenues for SZ science include the measurement and characterisation of low density regions of our Universe such as filaments between galaxy clusters \citep{deGraaff:2017byg,Tanimura:2017ixt} and cosmic voids \citep{Alonso:2017jaf}.
\end{itemize}

\vspace{1mm}
{\small
\section*{Acknowledgments}
We are especially grateful to Eiichiro Komatsu as well as Colin Hill and Marilena LoVerde for numerous insightful discussions, and to Fabien Lacasa  and a referee for many suggestions on the first version of this manuscript. BB and JC thank the cosmology group at JBCA, and BB  also thanks  David Alonso,  Brad Benson, Ryu Makiya, Ian McCarthy  for useful interactions relevant to this project. Finally, BB and TB are grateful to Jean-Baptiste Melin, Eunseong Lee and in particular Florian Ruppin for discussions and help with the cluster counts likelihood. We also thank Antony Lewis for providing the publicly available GetDist package that we used for our contour plots. \\
\indent This work was supported in part by an ERC Consolidator Grant (CMBSPEC), No. 725456 and DOE DE-SC0017848 and the Deutsche Forschungsgemeinschaft through the graduate school “Particle and Astroparticle Physics in the Light of the LHC” and through the individual grant “Cosmological probes of dark matter properties”. Simulations for this work were performed with computing resources granted by the RWTH High Performance Computing cluster under project RWTH0113 and by JARA-HPC from RWTH Aachen University under JARA0184. 
JC was supported by the Royal Society as a Royal Society University Research Fellow at the University of Manchester, UK.
This analysis is based on observations obtained with Planck
(http://www.esa.int/Planck), an ESA science mission with instruments and
contributions directly funded by ESA Member States, NASA, and Canada.

\appendix
\section*{Appendix}
\renewcommand{\thesubsection}{\Alph{subsection}}

\subsection{Scaling of tSZ power spectrum with neutrino fraction}\label{ref:appendix-B}
In Section \ref{s:pheno} we presented the scaling of the amplitude of the tSZ power spectrum with the neutrino fraction $f_\nu\equiv \Omega_\nu/\Omega_\mathrm{m}$ in three different cases. In each case, a different  set of cosmological parameters are kept fixed as the neutrino mass varies. In general, the scaling of the tSZ power spectrum with cosmological parameters is given by $C_\ell^\mathrm{tSZ} \propto (\sigma_8^\mathrm{cb})^{8.1}\Omega_\mathrm{cb}^{3.2}B^{-3.2}h^{-1.7}$. The mass bias is assumed to be independent of the neutrino mass (it is related to the electron pressure within haloes and massive neutrinos should have a negligible effect or no effect at all on this parameter). Then, in all the three cases we can write $(\sigma_8^\mathrm{cb})^2 \propto(1-\alpha f_\nu)$, $\Omega_{\mathrm{cb}}\propto (1-\beta f_\nu)$ and $h\propto(1-\gamma f_\nu)$. At linear order in $f_\nu$, the scaling of the tSZ power spectrum reads
\begin{equation}
\frac{\Delta C_\ell^\mathrm{tSZ}}{C_\ell^\mathrm{tSZ}}=(-4.05\alpha - 3.2\beta +1.7\gamma)f_\nu.
\end{equation}
with $(\alpha,\beta,\gamma)=(5.5,1,0)$ in Case A and $(\alpha,\beta,\gamma)=(4.1,0,0)$ in Case C, as reported in Section \ref{s:pheno}. Using these numerical values we find the slopes $-25.5$ and $-16.6$ in Case A and C respectively. In Case B, the parameters $\Omega_\mathrm{b}h^2$ and  $\Omega_\mathrm{c}h^2$ are kept fixed. So we rewrite the scaling of the tSZ power spectrum as $C_\ell^\mathrm{tSZ} \propto (\sigma_8^\mathrm{cb})^{8.1}(\Omega_\mathrm{cb}h^2)^{3.2}B^{-3.2}h^{-8.1}$. At linear order in $f_\nu$, using $(\sigma_8^\mathrm{cb})^2 \propto(1-4.9 f_\nu)$ and $h\propto(1- 1.6 f_\nu)$ we find a slope of -6.88 for Case C.

\section*{Data Availability}

The data underlying this article are available in the public repositories of \verb|class_sz|\footnote{\url{https://github.com/borisbolliet/class\_sz}}  and \verb|MontePython|.\footnote{\url{https://github.com/brinckmann/montepython\_public}}

\bibliographystyle{mnras}
\bibliography{main}

\begin{thebibliography}{}
\makeatletter
\relax
\def\mn@urlcharsother{\let\do\@makeother \do\$\do\&\do\#\do\^\do\_\do\%\do\~}
\def\mn@doi{\begingroup\mn@urlcharsother \@ifnextchar [ {\mn@doi@}
  {\mn@doi@[]}}
\def\mn@doi@[#1]#2{\def\@tempa{#1}\ifx\@tempa\@empty \href
  {http://dx.doi.org/#2} {doi:#2}\else \href {http://dx.doi.org/#2} {#1}\fi
  \endgroup}
\def\mn@eprint#1#2{\mn@eprint@#1:#2::\@nil}
\def\mn@eprint@arXiv#1{\href {http://arxiv.org/abs/#1} {{\tt arXiv:#1}}}
\def\mn@eprint@dblp#1{\href {http://dblp.uni-trier.de/rec/bibtex/#1.xml}
  {dblp:#1}}
\def\mn@eprint@#1:#2:#3:#4\@nil{\def\@tempa {#1}\def\@tempb {#2}\def\@tempc
  {#3}\ifx \@tempc \@empty \let \@tempc \@tempb \let \@tempb \@tempa \fi \ifx
  \@tempb \@empty \def\@tempb {arXiv}\fi \@ifundefined
  {mn@eprint@\@tempb}{\@tempb:\@tempc}{\expandafter \expandafter \csname
  mn@eprint@\@tempb\endcsname \expandafter{\@tempc}}}

\bibitem[\protect\citeauthoryear{Abazajian et~al.}{Abazajian
  et~al.}{2016}]{Abazajian:2016yjj}
Abazajian K.~N.,  et~al., 2016, arXiv:1610.02743

\bibitem[\protect\citeauthoryear{Alam et~al.}{Alam et~al.}{2017}]{Alam:2016hwk}
Alam S.,  et~al., 2017, \mn@doi [Mon. Not. Roy. Astron. Soc.]
  {10.1093/mnras/stx721}, 470, 2617

\bibitem[\protect\citeauthoryear{Allison, Caucal, Calabrese, Dunkley  \&
  Louis}{Allison et~al.}{2015}]{Allison:2015qca}
Allison R.,  Caucal P.,  Calabrese E.,  Dunkley J.,   Louis T.,  2015, \mn@doi
  [Phys. Rev.] {10.1103/PhysRevD.92.123535}, D92, 123535

\bibitem[\protect\citeauthoryear{Alonso, Hill, Hlo{\v z}ek  \& Spergel}{Alonso
  et~al.}{2018}]{Alonso:2017jaf}
Alonso D.,  Hill J.~C.,  Hlo{\v z}ek R.,   Spergel D.~N.,  2018, \mn@doi [Phys.
  Rev.] {10.1103/PhysRevD.97.063514}, D97, 063514

\bibitem[\protect\citeauthoryear{Archidiacono, Brinckmann, Lesgourgues  \&
  Poulin}{Archidiacono et~al.}{2017}]{Archidiacono:2016lnv}
Archidiacono M.,  Brinckmann T.,  Lesgourgues J.,   Poulin V.,  2017, \mn@doi
  [JCAP] {10.1088/1475-7516/2017/02/052}, 1702, 052

\bibitem[\protect\citeauthoryear{{Arnaud}, {Pratt}, {Piffaretti},
  {B{\"o}hringer}, {Croston}  \& {Pointecouteau}}{{Arnaud}
  et~al.}{2010}]{2010A&A...517A..92A}
{Arnaud} M.,  {Pratt} G.~W.,  {Piffaretti} R.,  {B{\"o}hringer} H.,  {Croston}
  J.~H.,   {Pointecouteau} E.,  2010, \mn@doi [\aap]
  {10.1051/0004-6361/200913416}, \href
  {http://adsabs.harvard.edu/abs/2010A%26A...517A..92A} {517, A92}

\bibitem[\protect\citeauthoryear{{Audren}, {Lesgourgues}, {Benabed}  \&
  {Prunet}}{{Audren} et~al.}{2013a}]{2013JCAP...02..001A}
{Audren} B.,  {Lesgourgues} J.,  {Benabed} K.,   {Prunet} S.,  2013a, \mn@doi
  [JCAP] {10.1088/1475-7516/2013/02/001}, \href
  {http://adsabs.harvard.edu/abs/2013JCAP...02..001A} {2, 001}

\bibitem[\protect\citeauthoryear{Audren, Lesgourgues, Bird, Haehnelt  \&
  Viel}{Audren et~al.}{2013b}]{Audren:2012vy}
Audren B.,  Lesgourgues J.,  Bird S.,  Haehnelt M.~G.,   Viel M.,  2013b,
  \mn@doi [JCAP] {10.1088/1475-7516/2013/01/026}, 1301, 026

\bibitem[\protect\citeauthoryear{{Barbosa}, {Bartlett}, {Blanchard}  \&
  {Oukbir}}{{Barbosa} et~al.}{1996}]{Barbosa1996}
{Barbosa} D.,  {Bartlett} J.~G.,  {Blanchard} A.,   {Oukbir} J.,  1996, \aap,
  \href {https://ui.adsabs.harvard.edu/abs/1996A&A...314...13B} {314, 13}

\bibitem[\protect\citeauthoryear{Bashinsky \& Seljak}{Bashinsky \&
  Seljak}{2004}]{Bashinsky:2003tk}
Bashinsky S.,  Seljak U.,  2004, \mn@doi [Phys. Rev.]
  {10.1103/PhysRevD.69.083002}, D69, 083002

\bibitem[\protect\citeauthoryear{Basu et~al.}{Basu et~al.}{2019}]{Basu:2019rzm}
Basu K.,  et~al., 2019

\bibitem[\protect\citeauthoryear{{Battaglia}, {Bond}, {Pfrommer}  \&
  {Sievers}}{{Battaglia} et~al.}{2012}]{Battaglia2012}
{Battaglia} N.,  {Bond} J.~R.,  {Pfrommer} C.,   {Sievers} J.~L.,  2012,
  \mn@doi [\apj] {10.1088/0004-637X/758/2/74}, \href
  {https://ui.adsabs.harvard.edu/abs/2012ApJ...758...74B} {758, 74}

\bibitem[\protect\citeauthoryear{{Battye} \& {Weller}}{{Battye} \&
  {Weller}}{2003}]{Battye2003}
{Battye} R.~A.,  {Weller} J.,  2003, \mn@doi [\prd]
  {10.1103/PhysRevD.68.083506}, \href
  {https://ui.adsabs.harvard.edu/abs/2003PhRvD..68h3506B} {68, 083506}

\bibitem[\protect\citeauthoryear{Beutler et~al.,}{Beutler
  et~al.}{2011}]{Beutler:2011hx}
Beutler F.,  et~al., 2011, \mn@doi [Mon. Not. Roy. Astron. Soc.]
  {10.1111/j.1365-2966.2011.19250.x}, 416, 3017

\bibitem[\protect\citeauthoryear{Birkinshaw, Gull  \& Hardebeck}{Birkinshaw
  et~al.}{1984}]{Birkinshaw:1984aa}
Birkinshaw M.,  Gull S.~F.,   Hardebeck H.,  1984, Nature, 309, 34 EP

\bibitem[\protect\citeauthoryear{{Blas}, {Lesgourgues}  \& {Tram}}{{Blas}
  et~al.}{2011}]{2011JCAP...07..034B}
{Blas} D.,  {Lesgourgues} J.,   {Tram} T.,  2011, \mn@doi [JCAP]
  {10.1088/1475-7516/2011/07/034}, \href
  {http://adsabs.harvard.edu/abs/2011JCAP...07..034B} {7, 034}

\bibitem[\protect\citeauthoryear{Bleem et~al.}{Bleem
  et~al.}{2015}]{Bleem:2014iim}
Bleem L.~E.,  et~al., 2015, \mn@doi [Astrophys. J. Suppl.]
  {10.1088/0067-0049/216/2/27}, 216, 27

\bibitem[\protect\citeauthoryear{Bocquet, Saro, Dolag  \& Mohr}{Bocquet
  et~al.}{2016}]{Bocquet:2015pva}
Bocquet S.,  Saro A.,  Dolag K.,   Mohr J.~J.,  2016, \mn@doi [Mon. Not. Roy.
  Astron. Soc.] {10.1093/mnras/stv2657}, 456, 2361

\bibitem[\protect\citeauthoryear{Bocquet et~al.}{Bocquet
  et~al.}{2018}]{Bocquet:2018ukq}
Bocquet S.,  et~al., 2018, arXiv:1812.01679

\bibitem[\protect\citeauthoryear{Bolliet, Comis, Komatsu  \&
  Mac{\'\i}as-P{\'e}rez}{Bolliet et~al.}{2018}]{Bolliet:2017lha}
Bolliet B.,  Comis B.,  Komatsu E.,   Mac{\'\i}as-P{\'e}rez J.~F.,  2018,
  \mn@doi [Monthly Notices of the Royal Astronomical Society]
  {10.1093/mnras/sty823}, 477, 4957

\bibitem[\protect\citeauthoryear{Boyle}{Boyle}{2019}]{Boyle:2018rva}
Boyle A.,  2019, \mn@doi [JCAP] {10.1088/1475-7516/2019/04/038}, 1904, 038

\bibitem[\protect\citeauthoryear{Boyle \& Komatsu}{Boyle \&
  Komatsu}{2018}]{Boyle:2017lzt}
Boyle A.,  Komatsu E.,  2018, \mn@doi [JCAP] {10.1088/1475-7516/2018/03/035},
  1803, 035

\bibitem[\protect\citeauthoryear{Brinckmann \& Lesgourgues}{Brinckmann \&
  Lesgourgues}{2018}]{Brinckmann:2018cvx}
Brinckmann T.,  Lesgourgues J.,  2018, arXiv:1804.07261

\bibitem[\protect\citeauthoryear{Brinckmann, Hooper, Archidiacono, Lesgourgues
  \& Sprenger}{Brinckmann et~al.}{2019}]{Brinckmann:2018owf}
Brinckmann T.,  Hooper D.~C.,  Archidiacono M.,  Lesgourgues J.,   Sprenger T.,
   2019, \mn@doi [JCAP] {10.1088/1475-7516/2019/01/059}, 1901, 059

\bibitem[\protect\citeauthoryear{Buen-Abad, Schmaltz, Lesgourgues  \&
  Brinckmann}{Buen-Abad et~al.}{2018}]{Buen-Abad:2017gxg}
Buen-Abad M.~A.,  Schmaltz M.,  Lesgourgues J.,   Brinckmann T.,  2018, \mn@doi
  [JCAP] {10.1088/1475-7516/2018/01/008}, 1801, 008

\bibitem[\protect\citeauthoryear{Burenin, Vikhlinin, Hornstrup, Ebeling,
  Quintana  \& Mescheryakov}{Burenin et~al.}{2007}]{Burenin:2006td}
Burenin R.~A.,  Vikhlinin A.,  Hornstrup A.,  Ebeling H.,  Quintana H.,
  Mescheryakov A.,  2007, \mn@doi [Astrophys. J. Suppl.] {10.1086/519457}, 172,
  561

\bibitem[\protect\citeauthoryear{Calabrese, Alonso  \& Dunkley}{Calabrese
  et~al.}{2017a}]{Calabrese:2016eii}
Calabrese E.,  Alonso D.,   Dunkley J.,  2017a, \mn@doi [Phys. Rev.]
  {10.1103/PhysRevD.95.063504}, D95, 063504

\bibitem[\protect\citeauthoryear{Calabrese et~al.}{Calabrese
  et~al.}{2017b}]{Calabrese:2017ypx}
Calabrese E.,  et~al., 2017b, \mn@doi [Phys. Rev.]
  {10.1103/PhysRevD.95.063525}, D95, 063525

\bibitem[\protect\citeauthoryear{Carlstrom, Holder  \& Reese}{Carlstrom
  et~al.}{2002}]{Carlstrom:2002na}
Carlstrom J.~E.,  Holder G.~P.,   Reese E.~D.,  2002, \mn@doi [Ann. Rev.
  Astron. Astrophys.] {10.1146/annurev.astro.40.060401.093803}, 40, 643

\bibitem[\protect\citeauthoryear{Castorina, Sefusatti, Sheth,
  Villaescusa-Navarro  \& Viel}{Castorina et~al.}{2014}]{Castorina:2013wga}
Castorina E.,  Sefusatti E.,  Sheth R.~K.,  Villaescusa-Navarro F.,   Viel M.,
  2014, \mn@doi [JCAP] {10.1088/1475-7516/2014/02/049}, 02, 049

\bibitem[\protect\citeauthoryear{Cooray \& Sheth}{Cooray \&
  Sheth}{2002}]{Cooray:2002dia}
Cooray A.,  Sheth R.~K.,  2002, \mn@doi [Phys. Rept.]
  {10.1016/S0370-1573(02)00276-4}, 372, 1

\bibitem[\protect\citeauthoryear{Costanzi, Villaescusa-Navarro, Viel, Xia,
  Borgani, Castorina  \& Sefusatti}{Costanzi
  et~al.}{2013}]{Costanzi:2013bhatex}
Costanzi M.,  Villaescusa-Navarro F.,  Viel M.,  Xia J.-Q.,  Borgani S.,
  Castorina E.,   Sefusatti E.,  2013, \mn@doi [JCAP]
  {10.1088/1475-7516/2013/12/012}, 1312, 012

\bibitem[\protect\citeauthoryear{Cuesta, Niro  \& Verde}{Cuesta
  et~al.}{2016}]{Cuesta:2015iho}
Cuesta A.~J.,  Niro V.,   Verde L.,  2016, \mn@doi [Phys. Dark Univ.]
  {10.1016/j.dark.2016.04.005}, 13, 77

\bibitem[\protect\citeauthoryear{{DES Collaboration}}{{DES
  Collaboration}}{2018}]{Abbott:2017wau}
{DES Collaboration} 2018, \mn@doi [Phys. Rev.] {10.1103/PhysRevD.98.043526},
  D98, 043526

\bibitem[\protect\citeauthoryear{Di~Valentino et~al.}{Di~Valentino
  et~al.}{2018}]{DiValentino:2016foa}
Di~Valentino E.,  et~al., 2018, \mn@doi [JCAP] {10.1088/1475-7516/2018/04/017},
  1804, 017

\bibitem[\protect\citeauthoryear{Dolag, Komatsu  \& Sunyaev}{Dolag
  et~al.}{2016}]{Dolag:2015dta}
Dolag K.,  Komatsu E.,   Sunyaev R.,  2016, \mn@doi [Mon. Not. Roy. Astron.
  Soc.] {10.1093/mnras/stw2035}, 463, 1797

\bibitem[\protect\citeauthoryear{Efstathiou \& Migliaccio}{Efstathiou \&
  Migliaccio}{2012}]{doi:10.1111/j.1365-2966.2012.21059.x}
Efstathiou G.,  Migliaccio M.,  2012, \mn@doi [Monthly Notices of the Royal
  Astronomical Society] {10.1111/j.1365-2966.2012.21059.x}, 423, 2492

\bibitem[\protect\citeauthoryear{Eisenstein \& Hu}{Eisenstein \&
  Hu}{1997}]{Eisenstein:1997jh}
Eisenstein D.~J.,  Hu W.,  1997, \mn@doi [Astrophys. J.] {10.1086/306640}, 511,
  5

\bibitem[\protect\citeauthoryear{Esteban, Gonzalez-Garcia, Hernandez-Cabezudo,
  Maltoni  \& Schwetz}{Esteban et~al.}{2019}]{Esteban:2018azc}
Esteban I.,  Gonzalez-Garcia M.~C.,  Hernandez-Cabezudo A.,  Maltoni M.,
  Schwetz T.,  2019, \mn@doi [JHEP] {10.1007/JHEP01(2019)106}, 01, 106

\bibitem[\protect\citeauthoryear{Font-Ribera, McDonald, Mostek, Reid, Seo  \&
  Slosar}{Font-Ribera et~al.}{2014}]{Font-Ribera:2013rwa}
Font-Ribera A.,  McDonald P.,  Mostek N.,  Reid B.~A.,  Seo H.-J.,   Slosar A.,
   2014, \mn@doi [JCAP] {10.1088/1475-7516/2014/05/023}, 1405, 023

\bibitem[\protect\citeauthoryear{George et~al.}{George
  et~al.}{2015}]{George:2014oba}
George E.~M.,  et~al., 2015, \mn@doi [Astrophys. J.]
  {10.1088/0004-637X/799/2/177}, 799, 177

\bibitem[\protect\citeauthoryear{Giusarma, Gerbino, Mena, Vagnozzi, Ho  \&
  Freese}{Giusarma et~al.}{2016}]{Giusarma:2016phn}
Giusarma E.,  Gerbino M.,  Mena O.,  Vagnozzi S.,  Ho S.,   Freese K.,  2016,
  \mn@doi [Phys. Rev.] {10.1103/PhysRevD.94.083522}, D94, 083522

\bibitem[\protect\citeauthoryear{Handley \& Lemos}{Handley \&
  Lemos}{2019}]{Handley:2019wlz}
Handley W.,  Lemos P.,  2019, arXiv:1902.04029

\bibitem[\protect\citeauthoryear{Hannestad}{Hannestad}{2010}]{Hannestad:2010kz}
Hannestad S.,  2010, \mn@doi [Prog. Part. Nucl. Phys.]
  {10.1016/j.ppnp.2010.07.001}, 65, 185

\bibitem[\protect\citeauthoryear{Hasselfield et~al.}{Hasselfield
  et~al.}{2013}]{Hasselfield:2013wf}
Hasselfield M.,  et~al., 2013, \mn@doi [JCAP] {10.1088/1475-7516/2013/07/008},
  7, 008

\bibitem[\protect\citeauthoryear{Henson, Barnes, Kay, McCarthy  \&
  Schaye}{Henson et~al.}{2017}]{Henson:2016eip}
Henson M.~A.,  Barnes D.~J.,  Kay S.~T.,  McCarthy I.~G.,   Schaye J.,  2017,
  \mn@doi [Mon. Not. Roy. Astron. Soc.] {10.1093/mnras/stw2899}, 465, 3361

\bibitem[\protect\citeauthoryear{Hildebrandt et~al.}{Hildebrandt
  et~al.}{2018}]{Hildebrandt:2018yau}
Hildebrandt H.,  et~al., 2018, arXiv:1812.06076

\bibitem[\protect\citeauthoryear{Hill \& Pajer}{Hill \&
  Pajer}{2013}]{Hill:2013baa}
Hill J.~C.,  Pajer E.,  2013, \mn@doi [Phys. Rev.]
  {10.1103/PhysRevD.88.063526}, D88, 063526

\bibitem[\protect\citeauthoryear{Hill et~al.}{Hill et~al.}{2014}]{Hill:2014pyr}
Hill J.~C.,  et~al., 2014, arXiv:1411.8004

\bibitem[\protect\citeauthoryear{Hilton et~al.}{Hilton
  et~al.}{2018}]{Hilton:2017gal}
Hilton M.,  et~al., 2018, \mn@doi [Astrophys. J. Suppl.]
  {10.3847/1538-4365/aaa6cb}, 235, 20

\bibitem[\protect\citeauthoryear{Hoekstra, Herbonnet, Muzzin, Babul, Mahdavi,
  Viola  \& Cacciato}{Hoekstra et~al.}{2015}]{Hoekstra:2015gda}
Hoekstra H.,  Herbonnet R.,  Muzzin A.,  Babul A.,  Mahdavi A.,  Viola M.,
  Cacciato M.,  2015, \mn@doi [Mon. Not. Roy. Astron. Soc.]
  {10.1093/mnras/stv275}, 449, 685

\bibitem[\protect\citeauthoryear{{Holder}, {Haiman}  \& {Mohr}}{{Holder}
  et~al.}{2001}]{Holder2001}
{Holder} G.,  {Haiman} Z.,   {Mohr} J.~J.,  2001, \mn@doi [\apj]
  {10.1086/324309}, \href
  {https://ui.adsabs.harvard.edu/abs/2001ApJ...560L.111H} {560, L111}

\bibitem[\protect\citeauthoryear{Horowitz \& Seljak}{Horowitz \&
  Seljak}{2017}]{Horowitz:2016dwk}
Horowitz B.,  Seljak U.,  2017, \mn@doi [Mon. Not. Roy. Astron. Soc.]
  {10.1093/mnras/stx766}, 469, 394

\bibitem[\protect\citeauthoryear{Hu \& Eisenstein}{Hu \&
  Eisenstein}{1998}]{Hu:1997vi}
Hu W.,  Eisenstein D.~J.,  1998, \mn@doi [Astrophys. J.] {10.1086/305585}, 498,
  497

\bibitem[\protect\citeauthoryear{Hu, Eisenstein  \& Tegmark}{Hu
  et~al.}{1998}]{Hu:1997mj}
Hu W.,  Eisenstein D.~J.,   Tegmark M.,  1998, \mn@doi [Phys. Rev. Lett.]
  {10.1103/PhysRevLett.80.5255}, 80, 5255

\bibitem[\protect\citeauthoryear{Hurier \& Lacasa}{Hurier \&
  Lacasa}{2017}]{Hurier:2017jgi}
Hurier G.,  Lacasa F.,  2017, \mn@doi [Astron. Astrophys.]
  {10.1051/0004-6361/201630041}, 604, A71

\bibitem[\protect\citeauthoryear{Ichiki \& Takada}{Ichiki \&
  Takada}{2012}]{Ichiki:2011ue}
Ichiki K.,  Takada M.,  2012, \mn@doi [Phys. Rev.]
  {10.1103/PhysRevD.85.063521}, D85, 063521

\bibitem[\protect\citeauthoryear{Ichiki, Yoo  \& Oguri}{Ichiki
  et~al.}{2016}]{Ichiki:2015gia}
Ichiki K.,  Yoo C.-M.,   Oguri M.,  2016, \mn@doi [Phys. Rev.]
  {10.1103/PhysRevD.93.023529}, D93, 023529

\bibitem[\protect\citeauthoryear{{Kaiser}}{{Kaiser}}{1986}]{1986MNRAS.222..323K}
{Kaiser} N.,  1986, \mn@doi [mnras] {10.1093/mnras/222.2.323}, \href
  {http://adsabs.harvard.edu/abs/1986MNRAS.222..323K} {222, 323}

\bibitem[\protect\citeauthoryear{{Kaiser}}{{Kaiser}}{1991}]{1991ApJ...383..104K}
{Kaiser} N.,  1991, \mn@doi [\apj] {10.1086/170768}, \href
  {https://ui.adsabs.harvard.edu/abs/1991ApJ...383..104K} {383, 104}

\bibitem[\protect\citeauthoryear{Kajita}{Kajita}{2016}]{Kajita:2016cak}
Kajita T.,  2016, \mn@doi [Rev. Mod. Phys.] {10.1103/RevModPhys.88.030501}, 88,
  030501

\bibitem[\protect\citeauthoryear{Kay, Thomas, Jenkins  \& Pearce}{Kay
  et~al.}{2004}]{Kay:2004wp}
Kay S.~T.,  Thomas P.~A.,  Jenkins A.,   Pearce F.~R.,  2004, \mn@doi [Mon.
  Not. Roy. Astron. Soc.] {10.1111/j.1365-2966.2004.08383.x}, 355, 1091

\bibitem[\protect\citeauthoryear{Kodwani, Alsono  \& Ferreira}{Kodwani
  et~al.}{2019}]{Kodwani_2019}
Kodwani D.,  Alsono D.,   Ferreira P.,  2019, \mn@doi [The Open Journal of
  Astrophysics] {10.21105/astro.1811.11584}, 2

\bibitem[\protect\citeauthoryear{Komatsu \& Kitayama}{Komatsu \&
  Kitayama}{1999}]{Komatsu:1999ev}
Komatsu E.,  Kitayama T.,  1999, \mn@doi [Astrophys. J.] {10.1086/312364}, 526,
  L1

\bibitem[\protect\citeauthoryear{Komatsu \& Seljak}{Komatsu \&
  Seljak}{2002}]{Komatsu:2002wc}
Komatsu E.,  Seljak U.,  2002, \mn@doi [Mon. Not. Roy. Astron. Soc.]
  {10.1046/j.1365-8711.2002.05889.x}, 336, 1256

\bibitem[\protect\citeauthoryear{{Kravtsov}, {Klypin}  \& {Hoffman}}{{Kravtsov}
  et~al.}{2002}]{2002ApJ...571..563K}
{Kravtsov} A.~V.,  {Klypin} A.,   {Hoffman} Y.,  2002, \mn@doi [\apj]
  {10.1086/340046}, \href {http://adsabs.harvard.edu/abs/2002ApJ...571..563K}
  {571, 563}

\bibitem[\protect\citeauthoryear{Kravtsov, Vikhlinin  \& Nagai}{Kravtsov
  et~al.}{2006}]{Kravtsov:2006db}
Kravtsov A.~V.,  Vikhlinin A.,   Nagai D.,  2006, \mn@doi [Astrophys. J.]
  {10.1086/506319}, 650, 128

\bibitem[\protect\citeauthoryear{Kreisch, Cyr-Racine  \& Dor{\'e}}{Kreisch
  et~al.}{2019}]{Kreisch:2019yzn}
Kreisch C.~D.,  Cyr-Racine F.-Y.,   Dor{\'e} O.,  2019, arXiv:1902.00534

\bibitem[\protect\citeauthoryear{Lattanzi \& Gerbino}{Lattanzi \&
  Gerbino}{2018}]{Lattanzi:2017ubx}
Lattanzi M.,  Gerbino M.,  2018, \mn@doi [Front.in Phys.]
  {10.3389/fphy.2017.00070}, 5, 70

\bibitem[\protect\citeauthoryear{Lau, Kravtsov  \& Nagai}{Lau
  et~al.}{2009}]{Lau:2009qm}
Lau E.~T.,  Kravtsov A.~V.,   Nagai D.,  2009, \mn@doi [Astrophys. J.]
  {10.1088/0004-637X/705/2/1129}, 705, 1129

\bibitem[\protect\citeauthoryear{Lesgourgues \& Pastor}{Lesgourgues \&
  Pastor}{2006}]{Lesgourgues:2006nd}
Lesgourgues J.,  Pastor S.,  2006, \mn@doi [Phys. Rept.]
  {10.1016/j.physrep.2006.04.001}, 429, 307

\bibitem[\protect\citeauthoryear{Lesgourgues, Mangano, Miele  \&
  Pastor}{Lesgourgues et~al.}{2013}]{Lesgourgues:2018ncw}
Lesgourgues J.,  Mangano G.,  Miele G.,   Pastor S.,  2013, {Neutrino
  Cosmology}.
Cambridge University Press

\bibitem[\protect\citeauthoryear{Lewis \& Bridle}{Lewis \&
  Bridle}{2002}]{Lewis:2002ah}
Lewis A.,  Bridle S.,  2002, \mn@doi [\prd] {10.1103/PhysRevD.66.103511}, 66,
  103511

\bibitem[\protect\citeauthoryear{Liu, Pritchard, Allison, Parsons, Seljak  \&
  Sherwin}{Liu et~al.}{2016}]{Liu:2015txa}
Liu A.,  Pritchard J.~R.,  Allison R.,  Parsons A.~R.,  Seljak U.,   Sherwin
  B.~D.,  2016, \mn@doi [Phys. Rev.] {10.1103/PhysRevD.93.043013}, D93, 043013

\bibitem[\protect\citeauthoryear{LoVerde}{LoVerde}{2014}]{LoVerde:2014rxa}
LoVerde M.,  2014, \mn@doi [Phys. Rev.] {10.1103/PhysRevD.90.083518}, D90,
  083518

\bibitem[\protect\citeauthoryear{LoVerde}{LoVerde}{2016}]{LoVerde:2016ahu}
LoVerde M.,  2016, \mn@doi [Phys. Rev.] {10.1103/PhysRevD.93.103526}, D93,
  103526

\bibitem[\protect\citeauthoryear{Lorenz, Calabrese  \& Alonso}{Lorenz
  et~al.}{2017}]{Lorenz:2017fgo}
Lorenz C.~S.,  Calabrese E.,   Alonso D.,  2017, \mn@doi [Phys. Rev.]
  {10.1103/PhysRevD.96.043510}, D96, 043510

\bibitem[\protect\citeauthoryear{Louis \& Alonso}{Louis \&
  Alonso}{2017}]{Louis:2016gvv}
Louis T.,  Alonso D.,  2017, \mn@doi [Phys. Rev.] {10.1103/PhysRevD.95.043517},
  D95, 043517

\bibitem[\protect\citeauthoryear{Lukic, Heitmann, Habib, Bashinsky  \&
  Ricker}{Lukic et~al.}{2007}]{Lukic:2007fc}
Lukic Z.,  Heitmann K.,  Habib S.,  Bashinsky S.,   Ricker P.~M.,  2007,
  \mn@doi [Astrophys. J.] {10.1086/523083}, 671, 1160

\bibitem[\protect\citeauthoryear{Makiya, Ando  \& Komatsu}{Makiya
  et~al.}{2018}]{Makiya:2018pda}
Makiya R.,  Ando S.,   Komatsu E.,  2018, \mn@doi [Monthly Notices of the Royal
  Astronomical Society] {10.1093/mnras/sty2031}, 480, 3928

\bibitem[\protect\citeauthoryear{Mangano, Miele, Pastor, Pinto, Pisanti  \&
  Serpico}{Mangano et~al.}{2005}]{Mangano:2005cc}
Mangano G.,  Miele G.,  Pastor S.,  Pinto T.,  Pisanti O.,   Serpico P.~D.,
  2005, \mn@doi [Nucl. Phys.] {10.1016/j.nuclphysb.2005.09.041}, B729, 221

\bibitem[\protect\citeauthoryear{Mccarthy, Bird, Schaye, Harnois-Deraps, Font
  \& Van~Waerbeke}{Mccarthy et~al.}{2018}]{McCarthy:2017csu}
Mccarthy I.~G.,  Bird S.,  Schaye J.,  Harnois-Deraps J.,  Font A.~S.,
  Van~Waerbeke L.,  2018, \mn@doi [Mon. Not. Roy. Astron. Soc.]
  {10.1093/mnras/sty377}, 476, 2999

\bibitem[\protect\citeauthoryear{Medezinski et~al.,}{Medezinski
  et~al.}{2017}]{HSC}
Medezinski E.,  et~al., 2017, \mn@doi [Publications of the Astronomical Society
  of Japan] {10.1093/pasj/psx128}, 70

\bibitem[\protect\citeauthoryear{Melin \& Bartlett}{Melin \&
  Bartlett}{2015}]{Melin:2014uaa}
Melin J.-B.,  Bartlett J.~G.,  2015, \mn@doi [Astron. Astrophys.]
  {10.1051/0004-6361/201424720}, 578, A21

\bibitem[\protect\citeauthoryear{Mishra-Sharma, Alonso  \&
  Dunkley}{Mishra-Sharma et~al.}{2018}]{Mishra-Sharma:2018ykh}
Mishra-Sharma S.,  Alonso D.,   Dunkley J.,  2018, \mn@doi [Phys. Rev.]
  {10.1103/PhysRevD.97.123544}, D97, 123544

\bibitem[\protect\citeauthoryear{Mroczkowski et~al.}{Mroczkowski
  et~al.}{2019}]{Mroczkowski:2018nrv}
Mroczkowski T.,  et~al., 2019, \mn@doi [Space Sci. Rev.]
  {10.1007/s11214-019-0581-2}, 215, 17

\bibitem[\protect\citeauthoryear{Nagai}{Nagai}{2006}]{Nagai:2005wx}
Nagai D.,  2006, \mn@doi [Astrophys. J.] {10.1086/506467}, 650, 538

\bibitem[\protect\citeauthoryear{{Nagai}, {Vikhlinin}  \& {Kravtsov}}{{Nagai}
  et~al.}{2007a}]{2007ApJ...655...98N}
{Nagai} D.,  {Vikhlinin} A.,   {Kravtsov} A.~V.,  2007a, \mn@doi [\apj]
  {10.1086/509868}, \href {http://adsabs.harvard.edu/abs/2007ApJ...655...98N}
  {655, 98}

\bibitem[\protect\citeauthoryear{{Nagai}, {Kravtsov}  \& {Vikhlinin}}{{Nagai}
  et~al.}{2007b}]{2007ApJ...668....1N}
{Nagai} D.,  {Kravtsov} A.~V.,   {Vikhlinin} A.,  2007b, \mn@doi [\apj]
  {10.1086/521328}, \href {http://adsabs.harvard.edu/abs/2007ApJ...668....1N}
  {668, 1}

\bibitem[\protect\citeauthoryear{Navarro, Frenk  \& White}{Navarro
  et~al.}{1996}]{Navarro:1995iw}
Navarro J.~F.,  Frenk C.~S.,   White S. D.~M.,  1996, \mn@doi [Astrophys. J.]
  {10.1086/177173}, 462, 563

\bibitem[\protect\citeauthoryear{Nelson, Lau, Nagai, Rudd  \& Yu}{Nelson
  et~al.}{2014}]{Nelson:2013hwa}
Nelson K.,  Lau E.~T.,  Nagai D.,  Rudd D.~H.,   Yu L.,  2014, \mn@doi
  [Astrophys. J.] {10.1088/0004-637X/782/2/107}, 782, 107

\bibitem[\protect\citeauthoryear{Obuljen, Castorina, Villaescusa-Navarro  \&
  Viel}{Obuljen et~al.}{2018}]{Obuljen:2017jiy}
Obuljen A.,  Castorina E.,  Villaescusa-Navarro F.,   Viel M.,  2018, \mn@doi
  [JCAP] {10.1088/1475-7516/2018/05/004}, 1805, 004

\bibitem[\protect\citeauthoryear{Oldengott, Barenboim, Kahlen, Salvado  \&
  Schwarz}{Oldengott et~al.}{2019}]{Oldengott:2019lke}
Oldengott I.~M.,  Barenboim G.,  Kahlen S.,  Salvado J.,   Schwarz D.~J.,
  2019, \mn@doi [JCAP] {10.1088/1475-7516/2019/04/049}, 1904, 049

\bibitem[\protect\citeauthoryear{Palanque-Delabrouille
  et~al.}{Palanque-Delabrouille et~al.}{2015}]{Palanque-Delabrouille:2015pga}
Palanque-Delabrouille N.,  et~al., 2015, \mn@doi [JCAP]
  {10.1088/1475-7516/2015/11/011}, 1511, 011

\bibitem[\protect\citeauthoryear{{Peebles}, {Daly}  \& {Juszkiewicz}}{{Peebles}
  et~al.}{1989}]{1989ApJ...347..563P}
{Peebles} P.~J.~E.,  {Daly} R.~A.,   {Juszkiewicz} R.,  1989, \mn@doi [\apj]
  {10.1086/168149}, \href
  {https://ui.adsabs.harvard.edu/abs/1989ApJ...347..563P} {347, 563}

\bibitem[\protect\citeauthoryear{Perotto, Lesgourgues, Hannestad, Tu  \&
  Wong}{Perotto et~al.}{2006}]{Perotto:2006rj}
Perotto L.,  Lesgourgues J.,  Hannestad S.,  Tu H.,   Wong Y. Y.~Y.,  2006,
  \mn@doi [JCAP] {10.1088/1475-7516/2006/10/013}, 0610, 013

\bibitem[\protect\citeauthoryear{{Planck Collaboration}}{{Planck
  Collaboration}}{2013a}]{2013A&A...550A.131P}
{Planck Collaboration} 2013a, \mn@doi [Astron. Astrophys.]
  {10.1051/0004-6361/201220040}, \href
  {http://adsabs.harvard.edu/abs/2013A%26A...550A.131P} {550, A131}

\bibitem[\protect\citeauthoryear{{Planck Collaboration}}{{Planck
  Collaboration}}{2013b}]{Ade:2013qta}
{Planck Collaboration} 2013b, \mn@doi [Astron. Astrophys.]
  {10.1051/0004-6361/201321522}, 571, A21

\bibitem[\protect\citeauthoryear{{Planck Collaboration}}{{Planck
  Collaboration}}{2014a}]{Ade:2013kta}
{Planck Collaboration} 2014a, \mn@doi [Astron. Astrophys.]
  {10.1051/0004-6361/201321573}, 571, A15

\bibitem[\protect\citeauthoryear{{Planck Collaboration}}{{Planck
  Collaboration}}{2014b}]{Ade:2013lmv}
{Planck Collaboration} 2014b, \mn@doi [Astron. Astrophys.]
  {10.1051/0004-6361/201321521}, 571, A20

\bibitem[\protect\citeauthoryear{{Planck Collaboration}}{{Planck
  Collaboration}}{2015}]{Aghanim:2015eva}
{Planck Collaboration} 2015, \mn@doi [Astron. Astrophys.]
  {10.1051/0004-6361/201525826}, 594, A22

\bibitem[\protect\citeauthoryear{{Planck Collaboration}}{{Planck
  Collaboration}}{2016a}]{Aghanim:2015xee}
{Planck Collaboration} 2016a, \mn@doi [Astron. Astrophys.]
  {10.1051/0004-6361/201526926}, 594, A11

\bibitem[\protect\citeauthoryear{{Planck Collaboration}}{{Planck
  Collaboration}}{2016b}]{Ade:2015xua}
{Planck Collaboration} 2016b, \mn@doi [Astron. Astrophys.]
  {10.1051/0004-6361/201525830}, 594, A13

\bibitem[\protect\citeauthoryear{{Planck Collaboration}}{{Planck
  Collaboration}}{2016c}]{Ade:2015fva}
{Planck Collaboration} 2016c, \mn@doi [Astron. Astrophys.]
  {10.1051/0004-6361/201525833}, 594, A24

\bibitem[\protect\citeauthoryear{{Planck Collaboration}}{{Planck
  Collaboration}}{2016d}]{Ade:2015gva}
{Planck Collaboration} 2016d, \mn@doi [Astron. Astrophys.]
  {10.1051/0004-6361/201525823}, 594, A27

\bibitem[\protect\citeauthoryear{{Planck Collaboration}}{{Planck
  Collaboration}}{2016e}]{Aghanim:2016yuo}
{Planck Collaboration} 2016e, \mn@doi [Astron. Astrophys.]
  {10.1051/0004-6361/201628890}, 596, A107

\bibitem[\protect\citeauthoryear{{Planck Collaboration}}{{Planck
  Collaboration}}{2018}]{Aghanim:2018eyx}
{Planck Collaboration} 2018

\bibitem[\protect\citeauthoryear{Pratt, Arnaud, Biviano, Eckert, Ettori, Nagai,
  Okabe  \& Reiprich}{Pratt et~al.}{2019}]{Pratt:2019cnf}
Pratt G.~W.,  Arnaud M.,  Biviano A.,  Eckert D.,  Ettori S.,  Nagai D.,  Okabe
  N.,   Reiprich T.~H.,  2019, \mn@doi [Space Sci. Rev.]
  {10.1007/s11214-019-0591-0}, 215, 25

\bibitem[\protect\citeauthoryear{{Reichardt} et~al.,}{{Reichardt}
  et~al.}{2012}]{2012ApJ...755...70R}
{Reichardt} C.~L.,  et~al., 2012, \mn@doi [\apj] {10.1088/0004-637X/755/1/70},
  \href {https://ui.adsabs.harvard.edu/abs/2012ApJ...755...70R} {755, 70}

\bibitem[\protect\citeauthoryear{Remazeilles, Bolliet, Rotti  \&
  Chluba}{Remazeilles et~al.}{2019}]{Remazeilles:2018laq}
Remazeilles M.,  Bolliet B.,  Rotti A.,   Chluba J.,  2019, \mn@doi [Mon. Not.
  Roy. Astron. Soc.] {10.1093/mnras/sty3352}, 483, 3459

\bibitem[\protect\citeauthoryear{Riess, Casertano, Yuan, Macri  \&
  Scolnic}{Riess et~al.}{2019}]{Riess:2019cxk}
Riess A.~G.,  Casertano S.,  Yuan W.,  Macri L.~M.,   Scolnic D.,  2019

\bibitem[\protect\citeauthoryear{Roncarelli, Carbone  \& Moscardini}{Roncarelli
  et~al.}{2015}]{Roncarelli:2014jla}
Roncarelli M.,  Carbone C.,   Moscardini L.,  2015, \mn@doi [Mon. Not. Roy.
  Astron. Soc.] {10.1093/mnras/stu2546}, 447, 1761

\bibitem[\protect\citeauthoryear{Ross, Samushia, Howlett, Percival, Burden  \&
  Manera}{Ross et~al.}{2015}]{Ross:2014qpa}
Ross A.~J.,  Samushia L.,  Howlett C.,  Percival W.~J.,  Burden A.,   Manera
  M.,  2015, \mn@doi [Mon. Not. Roy. Astron. Soc.] {10.1093/mnras/stv154}, 449,
  835

\bibitem[\protect\citeauthoryear{{Ruppin} et~al.,}{{Ruppin}
  et~al.}{2018}]{2018A&A...615A.112R}
{Ruppin} F.,  et~al., 2018, \mn@doi [\aap] {10.1051/0004-6361/201732558}, \href
  {https://ui.adsabs.harvard.edu/abs/2018A&A...615A.112R} {615, A112}

\bibitem[\protect\citeauthoryear{Ruppin, Mayet, Mac{\'\i}as-P{\'e}rez  \&
  Perotto}{Ruppin et~al.}{2019}]{Ruppin:2019deo}
Ruppin F.,  Mayet F.,  Mac{\'\i}as-P{\'e}rez J.~F.,   Perotto L.,  2019,
  \mn@doi [Monthly Notices of the Royal Astronomical Society]
  {10.1093/mnras/stz2669}, 490, 784

\bibitem[\protect\citeauthoryear{Salvati, Douspis  \& Aghanim}{Salvati
  et~al.}{2018}]{Salvati:2017rsn}
Salvati L.,  Douspis M.,   Aghanim N.,  2018, \mn@doi [Astron. Astrophys.]
  {10.1051/0004-6361/201731990}, 614, A13

\bibitem[\protect\citeauthoryear{Salvati, Douspis, Ritz, Aghanim  \&
  Babul}{Salvati et~al.}{2019}]{Salvati:2019zdp}
Salvati L.,  Douspis M.,  Ritz A.,  Aghanim N.,   Babul A.,  2019, \mn@doi
  [Astronomy & Astrophysics] {10.1051/0004-6361/201935041}, 626, A27

\bibitem[\protect\citeauthoryear{Sarazin}{Sarazin}{1986}]{RevModPhys.58.1}
Sarazin C.~L.,  1986, \mn@doi [Rev. Mod. Phys.] {10.1103/RevModPhys.58.1}, 58,
  1

\bibitem[\protect\citeauthoryear{Shi \& Komatsu}{Shi \&
  Komatsu}{2014}]{Shi:2014msa}
Shi X.,  Komatsu E.,  2014, \mn@doi [Mon. Not. Roy. Astron. Soc.]
  {10.1093/mnras/stu858}, 442, 521

\bibitem[\protect\citeauthoryear{Shi, Komatsu, Nelson  \& Nagai}{Shi
  et~al.}{2015}]{Shi:2014lua}
Shi X.,  Komatsu E.,  Nelson K.,   Nagai D.,  2015, \mn@doi [Mon. Not. Roy.
  Astron. Soc.] {10.1093/mnras/stv036, 10.1093/MNRASJ/stv036}, 448, 1020

\bibitem[\protect\citeauthoryear{Shi, Komatsu, Nagai  \& Lau}{Shi
  et~al.}{2016}]{Shi:2015fua}
Shi X.,  Komatsu E.,  Nagai D.,   Lau E.~T.,  2016, \mn@doi [Mon. Not. Roy.
  Astron. Soc.] {10.1093/mnras/stv2504}, 455, 2936

\bibitem[\protect\citeauthoryear{{Simons Collaboration}}{{Simons
  Collaboration}}{2019}]{Ade:2018sbj}
{Simons Collaboration} 2019, \mn@doi [Journal of Cosmology and Astroparticle
  Physics] {10.1088/1475-7516/2019/02/056}, 2019, 056

\bibitem[\protect\citeauthoryear{Smith et~al.}{Smith
  et~al.}{2016}]{Smith:2015qhs}
Smith G.~P.,  et~al., 2016, \mn@doi [Mon. Not. Roy. Astron. Soc.]
  {10.1093/mnrasl/slv175}, 456, L74

\bibitem[\protect\citeauthoryear{Sprenger, Archidiacono, Brinckmann, Clesse  \&
  Lesgourgues}{Sprenger et~al.}{2019}]{Sprenger:2018tdb}
Sprenger T.,  Archidiacono M.,  Brinckmann T.,  Clesse S.,   Lesgourgues J.,
  2019, \mn@doi [JCAP] {10.1088/1475-7516/2019/02/047}, 1902, 047

\bibitem[\protect\citeauthoryear{{Sunyaev} \& {Zeldovich}}{{Sunyaev} \&
  {Zeldovich}}{1972}]{1972CoASP...4..173S}
{Sunyaev} R.~A.,  {Zeldovich} Y.~B.,  1972, Comments on Astrophysics and Space
  Physics, \href {http://adsabs.harvard.edu/abs/1972CoASP...4..173S} {4, 173}

\bibitem[\protect\citeauthoryear{Tanimura et~al.,}{Tanimura
  et~al.}{2019}]{Tanimura:2017ixt}
Tanimura H.,  et~al., 2019, \mn@doi [Mon. Not. Roy. Astron. Soc.]
  {10.1093/mnras/sty3118}, 483, 223

\bibitem[\protect\citeauthoryear{Tinker, Kravtsov, Klypin, Abazajian, Warren,
  Yepes, Gottlober  \& Holz}{Tinker et~al.}{2008}]{Tinker:2008ff}
Tinker J.~L.,  Kravtsov A.~V.,  Klypin A.,  Abazajian K.,  Warren M.~S.,  Yepes
  G.,  Gottlober S.,   Holz D.~E.,  2008, \mn@doi [Astrophys. J.]
  {10.1086/591439}, 688, 709

\bibitem[\protect\citeauthoryear{Vagnozzi, Giusarma, Mena, Freese, Gerbino, Ho
  \& Lattanzi}{Vagnozzi et~al.}{2017}]{Vagnozzi:2017ovm}
Vagnozzi S.,  Giusarma E.,  Mena O.,  Freese K.,  Gerbino M.,  Ho S.,
  Lattanzi M.,  2017, \mn@doi [Phys. Rev.] {10.1103/PhysRevD.96.123503}, D96,
  123503

\bibitem[\protect\citeauthoryear{Vagnozzi, Brinckmann, Archidiacono, Freese,
  Gerbino, Lesgourgues  \& Sprenger}{Vagnozzi et~al.}{2018}]{Vagnozzi:2018pwo}
Vagnozzi S.,  Brinckmann T.,  Archidiacono M.,  Freese K.,  Gerbino M.,
  Lesgourgues J.,   Sprenger T.,  2018, \mn@doi [JCAP]
  {10.1088/1475-7516/2018/09/001}, 1809, 001

\bibitem[\protect\citeauthoryear{Vikhlinin, Kravtsov, Forman, Jones,
  Markevitch, Murray  \& Van~Speybroeck}{Vikhlinin
  et~al.}{2006}]{Vikhlinin:2005mp}
Vikhlinin A.,  Kravtsov A.,  Forman W.,  Jones C.,  Markevitch M.,  Murray
  S.~S.,   Van~Speybroeck L.,  2006, \mn@doi [Astrophys. J.] {10.1086/500288},
  640, 691

\bibitem[\protect\citeauthoryear{Vikhlinin et~al.}{Vikhlinin
  et~al.}{2009}]{Vikhlinin:2008cd}
Vikhlinin A.,  et~al., 2009, \mn@doi [Astrophys. J.]
  {10.1088/0004-637X/692/2/1033}, 692, 1033

\bibitem[\protect\citeauthoryear{Villaescusa-Navarro, Bull  \&
  Viel}{Villaescusa-Navarro et~al.}{2015}]{Villaescusa-Navarro:2015cca}
Villaescusa-Navarro F.,  Bull P.,   Viel M.,  2015, \mn@doi [Astrophys. J.]
  {10.1088/0004-637X/814/2/146}, 814, 146

\bibitem[\protect\citeauthoryear{Watts et~al.}{Watts
  et~al.}{2018}]{Watts:2018etg}
Watts D.~J.,  et~al., 2018, \mn@doi [Astrophys. J.] {10.3847/1538-4357/aad283},
  863, 121

\bibitem[\protect\citeauthoryear{Weinberg, Mortonson, Eisenstein, Hirata, Riess
   \& Rozo}{Weinberg et~al.}{2013}]{Weinberg:2012es}
Weinberg D.~H.,  Mortonson M.~J.,  Eisenstein D.~J.,  Hirata C.,  Riess A.~G.,
   Rozo E.,  2013, \mn@doi [Phys. Rept.] {10.1016/j.physrep.2013.05.001}, 530,
  87

\bibitem[\protect\citeauthoryear{{White}, {Efstathiou}  \& {Frenk}}{{White}
  et~al.}{1993}]{1993MNRAS.262.1023W}
{White} S.~D.~M.,  {Efstathiou} G.,   {Frenk} C.~S.,  1993, \mn@doi [\mnras]
  {10.1093/mnras/262.4.1023}, \href
  {http://adsabs.harvard.edu/abs/1993MNRAS.262.1023W} {262, 1023}

\bibitem[\protect\citeauthoryear{Yu, Knight, Sherwin, Ferraro, Knox  \&
  Schmittfull}{Yu et~al.}{2018}]{Yu:2018tem}
Yu B.,  Knight R.~Z.,  Sherwin B.~D.,  Ferraro S.,  Knox L.,   Schmittfull M.,
  2018, arXiv:1809.02120

\bibitem[\protect\citeauthoryear{{Zeldovich} \& {Sunyaev}}{{Zeldovich} \&
  {Sunyaev}}{1969}]{1969Ap&SS...4..301Z}
{Zeldovich} Y.~B.,  {Sunyaev} R.~A.,  1969, \mn@doi [\apss]
  {10.1007/BF00661821}, \href
  {http://adsabs.harvard.edu/abs/1969Ap%26SS...4..301Z} {4, 301}

\bibitem[\protect\citeauthoryear{Zhang \& Sheth}{Zhang \&
  Sheth}{2007}]{Zhang:2007psa}
Zhang P.,  Sheth R.~K.,  2007, \mn@doi [Astrophys. J.] {10.1086/522913}, 671,
  14

\bibitem[\protect\citeauthoryear{Zubeldia \& Challinor}{Zubeldia \&
  Challinor}{2019}]{Zubeldia:2019brr}
Zubeldia {\'I}.,  Challinor A.,  2019, arXiv:1904.07887

\bibitem[\protect\citeauthoryear{de Graaff, Cai, Heymans  \& Peacock}{de~Graaff
  et~al.}{2019}]{deGraaff:2017byg}
de Graaff A.,  Cai Y.-C.,  Heymans C.,   Peacock J.~A.,  2019, \mn@doi [Astron.
  Astrophys.] {10.1051/0004-6361/201935159}, 624, A48

\bibitem[\protect\citeauthoryear{{de Haan} et~al.,}{{de Haan}
  et~al.}{2016}]{2016ApJ...832...95D}
{de Haan} T.,  et~al., 2016, \mn@doi [\apj] {10.3847/0004-637X/832/1/95}, \href
  {http://adsabs.harvard.edu/abs/2016ApJ...832...95D} {832, 95}

\bibitem[\protect\citeauthoryear{de Salas \& Pastor}{de~Salas \&
  Pastor}{2016}]{deSalas:2016ztq}
de Salas P.~F.,  Pastor S.,  2016, \mn@doi [JCAP]
  {10.1088/1475-7516/2016/07/051}, 1607, 051

\bibitem[\protect\citeauthoryear{de Salas, Forero, Ternes, Tortola  \&
  Valle}{de~Salas et~al.}{2018}]{deSalas:2017kay}
de Salas P.~F.,  Forero D.~V.,  Ternes C.~A.,  Tortola M.,   Valle J. W.~F.,
  2018, \mn@doi [Phys. Lett.] {10.1016/j.physletb.2018.06.019}, B782, 633

\bibitem[\protect\citeauthoryear{von~der Linden et~al.}{von~der Linden
  et~al.}{2014}]{vonderLinden:2014haa}
von~der Linden A.,  et~al., 2014, \mn@doi [Mon. Not. Roy. Astron. Soc.]
  {10.1093/mnras/stu1423}, 443, 1973

\makeatother
\end{thebibliography}
}

\end{document}